\documentclass{ieeeaccess}
   \usepackage{cite}
\usepackage{amsmath,amssymb,amsfonts}
\usepackage{algorithmic}
\usepackage{graphicx}
\graphicspath{ {./images/} }
\usepackage{CJK}
\usepackage{textcomp}
\usepackage{comment}
\usepackage{hyperref}
\usepackage{flushend}
\usepackage{xcolor, soul}
\sethlcolor{white}
\usepackage{array}
\usepackage{multirow}
\usepackage{mathabx}

\newcommand\MyBox[2]{
 \fbox{\lower0.75cm
  \vbox to 1cm{\vfil
   \hbox to 1cm{\hfil\parbox{1.4cm}{#1\\#2}\hfil}
   \vfil}%
 }%
}

\def\BibTeX{{\rm B\kern-.05em{\sc i\kern-.025em b}\kern-.08em
 T\kern-.1667em\lower.7ex\hbox{E}\kern-.125emX}}
\begin{document}
\history{Date of publication xxxx 00, 0000, date of current version xxxx 00, 0000.}
\doi{doi: 10.1109/ACCESS.2024.3401855}

\title{Towards Long Range Detection of Elephants Using Seismic Signals; A Geophone-Sensor Interface for Embedded Systems}

\author{\uppercase{Jaliya L. Wijayaraja}\authorrefmark{1,*},\IEEEmembership{Member, IEEE}
\uppercase{Janaka L. Wijekoon}\authorrefmark{3,4,*},\IEEEmembership{Senior Member, IEEE}, \uppercase{Malitha Wijesundara}\authorrefmark{1.*},\IEEEmembership{Member, IEEE}
\uppercase{and L. J. Mendis Wickramasinghe}\authorrefmark{2}
}

\address[1]{Sri Lanka Institute of Information Technology, New Kandy Rd, Malabe 10115, Sri Lanka}
%\address[2]{Sri Lanka Institute of Information Technology, New Kandy Rd, Malabe 10115, Sri Lanka}
%\address[3]{Faculty of Humanities \& Sciences, Sri Lanka Institute of Information Technology, New Kandy Rd, Malabe 10115, Sri Lanka}
\address[2]{Herpetological Foundation of Sri Lanka (HFS),
31/5, Alwis Town, Hendala, Wattala 11300, Sri Lanka}
\address [3]{Victorian Institute of Technology, Adelaide, South Australia}
\address[4]{Department of System Design Engineering, Keio University, Yokohama, Japan}

\tfootnote{The authors extend their sincere gratitude to Mr. Thilak Premakantha, \hl{The} Director General of the Department of National Zoological Gardens, the staff of the National Zoological Garden-Dehiwala, and the elephant orphanage, Pinnawala, for facilitating the conduct of field research experiments. Additionally, thanks are due to Mr. Samitha Vidhanaarachchi, Mr. Binura Ganegoda, Ms. Senuri Hettiarachchi, Ms. Hansi Hewavidana, and Ms. Chamodi Abisheka Pitumpe, who are staff members of Faculty of Computing, Sri Lanka Institute of Information Technology, for their support in field testing. Finally, the authors sincerely thank Dr. Rohan Pethiyagoda, Dr. Kasun Prasanga, and Dr. Shanaka P. Abeysiriwardhana for generously dedicating their time and expertise \hl{for proofreading} this paper. The project was carried out with financial support from the SLIIT Postgraduate Research Grant.}

\markboth
{Author \headeretal: Preparation of Papers for IEEE TRANSACTIONS and JOURNALS}
{Author \headeretal: Preparation of Papers for IEEE TRANSACTIONS and JOURNALS}

\corresp{Corresponding author: Jaliya L. Wijayaraja (e-mail: jaliya.wijayaraja@gmail.com).}

\begin{abstract}
The long-distance detection of the presence of elephants is pivotal to addressing \hl{the} human-elephant conflict. IoT-based solutions utilizing seismic signals originating from the movement of elephants are a novel approach to solving this problem. This study introduces an instrumentation system comprising a specially designed geophone-sensor interface for non-invasive, long-range elephant detection using seismic waves while minimizing the vulnerability of seismic signals to noise. The geophone-sensor interface involves a cascade array of an instrumentation amplifier, a second-order Butterworth filter for signal filtering, and a signal amplifier. The introduced geophone-sensor interface was tested under laboratory conditions, and then real-world experiments were carried out for tamed, partly tamed, and untamed elephants. The experimental results reveal that the system remains stable within the tested frequency range from 1 Hz to 1 kHz and the temperature range of 10° C to 40° C. The system successfully captured the seismic signals generated by the footfalls of elephants within a maximum detection range of 155.6 m, with an overall detection accuracy of 99.5\%.
\end{abstract}


\titlepgskip=-15pt

\maketitle

\section{Introduction}
\label{sec:introduction}

Human-elephant conflict (HEC) not only poses a substantial threat to the survival of elephants but also results in serious and often fatal harm to humans \cite{leimgruber2003fragmentation}. HEC is intensifying due to anthropogenic land usage and human settlement in elephant habitats, resulting in a mutual loss of life for both humans and elephants \cite{leimgruber2003fragmentation, shaffer2019human, prakash2020human, santiapillai2010assessment, jadhav2012elephant, thennakoon2017impact, choudhury2004human, dublin2004searching, kopke2021human}. Disturbingly, between 2010 and 2019, Sri Lanka reported a total of 14,516 HEC-related incidents, resulting in 807 human deaths and 2,631 elephant deaths \cite{prakash2020human}. Specifically in Sri Lanka, wild elephants cause more harm to humans \hl{than} any other wild animal \cite{thennakoon2017impact,kopke2021human}.

Addressing HEC has been receiving increasing attention not only among researchers \cite{shaffer2019human,nguyen2022integrating,enukwa2017human,perera2009human,sitati2006assessing,elvitigala2015towards} but also from both government and non-government institutions \cite{mishra2022first,slt2021}. Farmers, on the other hand, commonly adhere to traditional deterrents such as communal guarding, placing barriers on elephant routes, and using substances like chilli grease, and thunder-flashes as solutions for HEC \cite{sitati2006assessing,hedges2010reducing}. Regrettably, none of these methods seem to solve or substantially reduce HEC.

Among the different modern remedial solutions for HEC, electric fences stand out as the most practical solution \cite{wijesinghe2011electric, wijesekera2021modern}. However, given that electric fences prove effective only when elephants directly encounter them, real-time tracking of elephants and early identification of potential threats is infeasible with electric fences \cite{raman2003living}. According to \cite{wijesekera2021modern,cambron2014design,wijesundara2016design} tracking usually involves the use of radio collars fitted with tracking devices. However, as pointed out in \cite{wijesekera2021modern, wijesundara2016design}, fitting such collars on wild elephants involves a risk that limits the application of this method.

Notably, both tracking collars and electric fences are intrusive approaches that require the physical manipulation of elephants, potentially leading to harm. Therefore, passive and non-invasive detection methods are encouraged. Among such methods, namely visual \cite{premarathna2020mage,premarathna2020cnn,ravikumar2020layered,pemasinghe2023development,jothibasu2023improvement,mondal2023design}, acoustic \cite{zeppelzauer2013acoustic,bjorck2019automatic,prince2014surveillance,arya2016design,ranasinghe2023enhanced,ramasubramanian2022averting, deowan2022warning,swider2022passive,hedwig2019acoustic}, and infrasonic monitoring \cite{prince2014surveillance}, the use of seismic signals for passive elephant detection has proven to be highly effective in tropical conditions and uneven terrains \cite{wood2005using,anni2015elephant}.

Seismic waves can be effectively sensed by geophones, which are proven to be not only cost-effective but also accurate \cite{faber1997geophone,hou2020piezoelectric}. The significant ability of geophones to generate microvolt-range voltages for weak seismic signals makes it possible to identify long-range signals \cite{koc2013hardware, kafadar2020geophone}, albeit with susceptibility to various ambient noises. \hl{Consequently, aiming to improve the detection range and the accuracy of long-range elephant detection, this study is focused on designing a sophisticated geophone-sensor interface to read long-range seismic signals by reducing the impact of noise and amplifying the signals to process and detect elephant footfalls using the unique characteristics of elephants \footnote{\hl{This study was tested only for Elepha Maximus Maximus (Sri Lankan Elephant which is a distinct subspecies of the Asian Elephants)}.}}

\hl{The geophone-sensor interface utilizes AD620 monolithic instrumentation amplifiers to boost microvolt-range signals and employs a second-order low-pass Butterworth filter for anti-aliasing and noise-filtering. It also utilizes an LM324 signal amplifier to amplify the signal and achieve optimum voltage levels for the Analog-to-Digital Converter (ADC) of the embedded system. This is implemented with an AVR ATmega328p microcontroller, to determine elephant footfalls.. The geophone-sensor interface was first verified for its performance using software simulations.} Then, it was used for controlled field experiments in the National Zoological Gardens, Dehiwala, Sri Lanka; the Pinnawala Elephant Orphanage, Rambukkana, Sri Lanka; and Digampathaha, Sri Lanka, aiming to confirm the accurate acquisition of seismic signals for detecting elephant footfalls and to validate the detection range of the system. All the tests were video-recorded, and the classified elephant footfalls were correlated with the video evidence (the ground truth).

\section{Background Survey}
\label{sec:Background}
\hl{Studies presented in \mbox{\cite{o2000seismic,mortimer2018classifying,gunther2004seismic}} explain that elephants can generate seismic signals through vocalization and locomotion, and that was the key to detecting elephants by using geophones. Particularly, in \mbox{\cite{o2000seismic}}, O’Connell-Rodwell et al. identified several key properties of seismic waves generated by elephants.} The footfall signature of the elephant (i.e., stomp) has a mean frequency of 24.03 Hz ($\pm$2.98), and the vocal pattern of elephants (i.e, rumble) has a mean of 20.66 Hz ($\pm$2.07). The authors further presented a mathematical model suggesting that the rumbles can be detected instrumentally above ambient noise at a distance of up to 16 km, while the stomps can be detected at a distance of 32 km.

\hl{
Following the above three studies, a limited number of studies have tested the use of geophones to capture seismic signals from elephants, and they can be categorized into three distinct groups according to the methods used for recording the seismic signals: utilizing a sound card as an interface, employing a standard seismic data acquisition system, or developing a custom embedded system. A summary of these studies is detailed in Table \mbox{\ref{tab:literature}}.}

\begin{table*}[!t]
 \centering
 \caption{\hl{\textbf{Comparison with previous studies (So- used a sound card, St- standard seismic data acquisition system, Em- incorporated generic embedded system, V- vocalization, L- locomotion Cl-close/along the Path, NA- not available/ not relevant }}}
 \begin{tabular}{|p{0.7 cm}|p{0.8 cm}|p{1cm}|p{1.2cm}|p{1.5 cm}|p{1.1 cm}|p{0.9cm}|p{1.4cm}|p{1.8 cm}|p{1cm}|p{1.6cm}|}
 \hline
 Ref.& Method Used & Elephant Type  & Seismic Event & Geophone Frequency & Geophone Axis  & Gain Value & Incorporated Filters & Sampling Frequency & Detection Range &  Reported Accuracy \\
 \hline
 
    \cite{wood2005using} & So & African  & L & 4.5 Hz & Vertical & NA & Yes & 44.1 KHz  & 100 m & 82\%\\
    \cite{nakandala2014detecting} & So & Asian  & L & 4.5 Hz & Vertical & NA & Yes & 44.1 KHz  & Cl & 50\%\\
     
    \cite{o2000seismic} & St & Asian  & V, L & 10Hz & Vertical & NA & Yes & 8000 Samples/s & 40 m & NA\\
    
    \cite{szenicer2022seismic} & St & African  & V, L & NA &3 Axis & NA & NA & 200 Hz & 150 m & 73-90\%\\
    
    \cite{reinwald2021seismic} & St & African  & V & 0.03 - 100 Hz & NA & NA & Yes & 200 Hz & NA & NA \\
      
    \cite{parihar2021seismic} & Em & Asian  & L & 10 Hz & NA & NA & Yes & 1 KHz  & 40 m & 85.5-93.69\%\\
    \cite{fazil2018iot} & Em & NA  & L & NA & NA & NA & NA & NA & NA & NA\\
    \cite{fernando2020gaja} & Em & Asian  & L & 100Hz & NA & NA & NA & NA & 3-5m & 100\%\\
    \cite{zetterqvist2023elephant} & Em & African  & L & 10 Hz & NA & NA & Yes & 474 Hz & 40m & 99.9\%\\
     \cite{thilakarathne2022prototype} & Em & NA  & L & NA & NA & 72.8 dB & Yes & NA & NA & 93\%\\
    \cite{parihar2022variational} & Em & Asian  & L & NA & NA & Na & NA & 1 KHz & NA & 73\%\\
  \hline
 \end{tabular}
 \label{tab:literature}
\end{table*}

Wood et al.\mbox{\cite{wood2005using}} successfully utilized a geophone with a high-end audio interface to capture and classify seismic signals from large mammals, including elephants. Similarly, Nakandala et al. in \mbox{\cite{nakandala2014detecting}} followed the same method without a high-end sound card. \hl{As given in Table \mbox{\ref{tab:literature}}, even though both methods used the same parameters, Nakandala et al. \mbox{\cite{nakandala2014detecting}} reported a significant drop in accuracy, highlighting the importance of using high-end sound cards such as the Pocket v2 to achieve accurate results. However, it is important to note that expensive high-end sound cards are designed as computer peripherals assuming indoor usage and have very limited survivability in outdoor environments. These may not ensure long-term endurance in challenging environments, especially where HEC is prevalent.}

Reinwald et al. in \mbox{\cite{reinwald2021seismic}}, and Szenicer et al. in \mbox{\cite{szenicer2022seismic}}, utilized a Guralp 6TD seismometer to detect elephants, and \mbox{\cite{szenicer2022seismic}} reports the highest detection range (i.e., 150 m) for African elephants. \hl{However, the main limitation of that study is that the training and testing datasets of the classifier have been acquired using different seismic stations, which may introduced bias and inconsistencies into the experiments. Despite the wide detection range and the sufficient accuracy demonstrated by seismometers such as the Guralp 6TD, their high deployment cost limits their use for data acquisition in expansive HEC-prone areas, especially in developing countries. Additionally, such devices require specialized skills to operate, which further limits their widespread use.}

\hl{Compared to two methods: sound card and seismometer, for detecting elephant locomotion using seismic waves, the use of generic embedded systems is not only cost-effective and easier to implement in rough terrains but also customizable according to the requirement \mbox{\cite{parihar2021seismic,fernando2020gaja,zetterqvist2023elephant,parihar2022variational,thilakarathne2022prototype}}}. Among these studies, a successful approach was taken by Parihar et al. in \mbox{\cite{parihar2021seismic}}, utilizing a highly sensitive geophone (85.8 V/m/s) with a single-stage low-pass RC filter and 16-bit ADC, achieving a tested range of up to 40 m. \hl{However, there was a significant decrease in reported accuracy concerning distance (at 20-40 m range) due to high-frequency background noise. Similarly, \mbox{\cite{fazil2018iot}}, \mbox{\cite{parihar2022variational}}, and \mbox{\cite{fernando2020gaja}} present IoT-enabled systems for HEC using seismic signals, and \mbox{\cite{fernando2020gaja}} shows a successful study using seismic signals to detect elephants while employing an amplifier and a bandpass filter. Notably, the authors of all studies highlight the need for further improvement in analog signal processing for better accuracy.}

\hl{Moreover, as indicated in Table \mbox{\ref{tab:literature}}, despite all the studies \mbox{\cite{parihar2021seismic,fernando2020gaja,zetterqvist2023elephant,parihar2022variational,thilakarathne2022prototype}} concluding that the use of embedded systems together with a geophone demonstrated promising accuracies for non-invasive elephant detection, none of the studies reported a detection range beyond 40 m. Unfortunately, apart from \mbox{\cite{thilakarathne2022prototype}}, all the studies failed to provide details regarding the gains employed in their systems. Similarly, none of the studies discusses the impact of gain adjustments on the detection range, a crucial factor that hinders the implementation of long-range elephant detection.} Notably, there is no evidence provided on evaluating the performance in tropical terrains, which is also a critical consideration when studying HEC.

\hl{The highlighted limitations, i.e. the requirement for improved amplification under different gain settings and noise filtration to enhance the elephant detection range (especially over long distances)\mbox{\cite{parihar2021seismic,fernando2020gaja,fazil2018iot}}, the high cost of equipment and the use of sophisticated technologies \mbox{\cite{reinwald2021seismic,szenicer2022seismic}}, the lack of rigorous testing in tropical terrains with adequate parameters (for noise filtration and signal amplification gains)
, and the limitation elucidated when interfacing of geophone to an embedded system \mbox{\cite{parihar2021seismic,fernando2020gaja,fazil2018iot}}, incur the necessity for a sophisticated geophone-sensor interface to improve the accuracy of detecting elephants over long distances. Hence, this study proposes a geophone-sensor interface with the aim of amplifying signals by employing adjustable gains and eliminating seismic noise for effective and accurate long-range detection of elephants based on their characteristics in tropical terrains.}

\section{Methodology}

\begin{figure}[!t]
 \centering
 \includegraphics[width=0.45\textwidth]{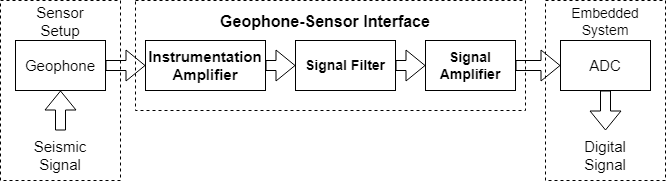}
 \caption{\textbf{Overall architecture of the geophone-sensor interface system}}
 \label{fig:process_diagram}
\end{figure}

Figure \ref{fig:process_diagram} illustrates the overall architecture of the instrumentation system. It comprises a sensor setup, the geophone-sensor interface, and an embedded system. The sensor setup consists of a geophone that primarily captures the seismic signals of elephants. The geophone-sensor interface gets signals from the sensor setup, amplifies, filters the noise, and delivers the signal to the embedded system. The embedded system utilizes an internal ADC and a microcontroller to identify elephant footfalls.

\subsection{Geophone selection and embedded system implementation}

\begin{figure}[!t]
 \centering
 \includegraphics[width=0.15\textwidth]{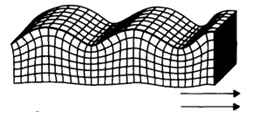}
 \caption{\textbf{Propagation of Rayleigh waves \cite{aicher1990vibrational}}}
 \label{fig:rayleigh_waves}
\end{figure}

\begin{figure}[!t]
 \centering
 \includegraphics[width=0.1\textwidth]{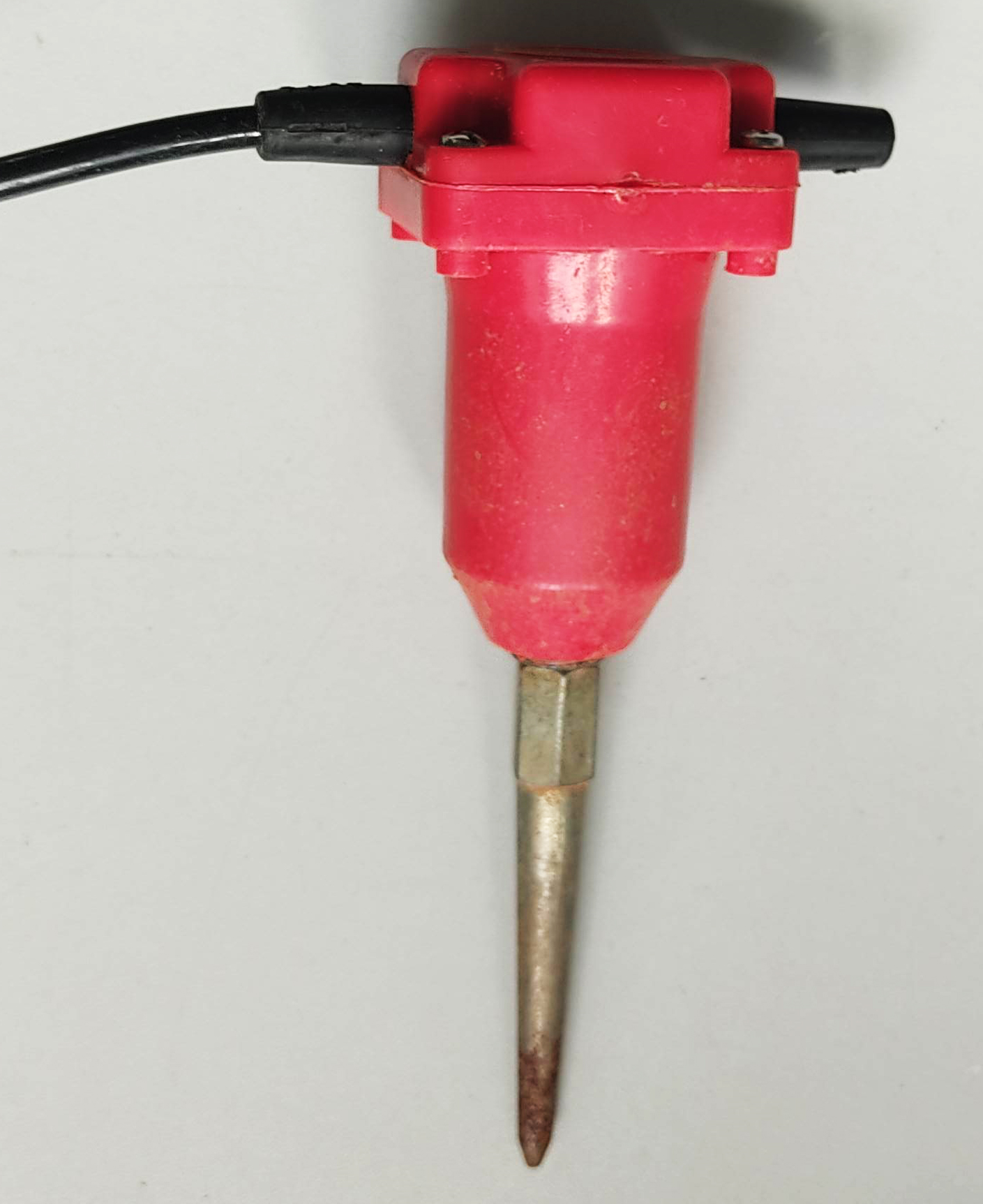}
 \caption{\textbf{RGI-HS10 Geophone}}
 \label{fig:geophone}
\end{figure}

\hl{The seismic waves generated by elephants are mostly Rayleigh waves and that contain a significantly higher energy.\mbox{\cite{o2007keeping,reuter1998elephant}}.} As illustrated in Fig. \ref{fig:rayleigh_waves}, Rayleigh waves are a type of surface waves that exhibits a vertical movement and diminishes exponentially with depth \cite{arnason2002properties,succi2000problems}. Therefore, a high sensitivity (RGI-HS 10) vertical geophone, with a sensitivity of 85.8 V/m/s, was selected for this study (see Fig. \ref{fig:geophone}) to precisely capture Rayleigh waves generated by elephants.

  \begin{figure*}[!t]
 \centering
 \includegraphics[width=0.55\textwidth]{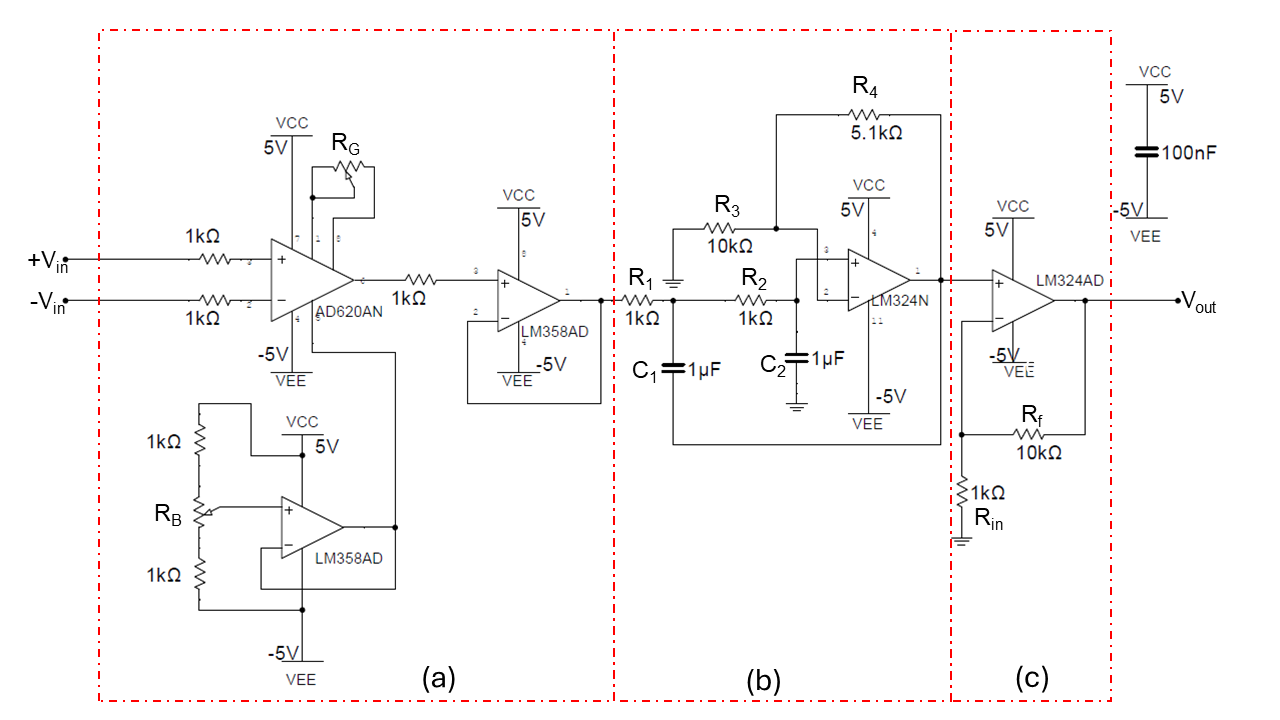}
 \caption{\hl{\textbf{Geophone-sensor interface circuit} }}
 \label{fig:circuit}
\end{figure*}

According to \cite{o2000seismic}, the RMS velocity of elephant rumbles is 0.437 µm/s at a distance of 3-12 m and is 0.264 µm/s at a distance of 40 m. For the corresponding distances, experimental peak-to-peak velocities for stomps were 48.7 µm/s and 17.0 µm/s, respectively. Based on the above facts, for the geophone: RGI-HS-10, output voltage $V_{out}$ was calculated using (\ref{eq:geophonesenitivity}), where $v_{seis}$ is the velocity of the seismic wave, and $S$ is the sensitivity of the geophone that is 85.8 V/m/s. The calculated voltages are presented in Table \ref{tab:voltages}.

\begin{equation}\label{eq:geophonesenitivity}
 V_{out}=v_{seis}\times S
\end{equation}

\begin{table}[!t]
 \centering
 \caption{\textbf{Calculated voltage outputs for identified seismic signal velocities of elephants}}
 \begin{tabular}{|c|c|c|}
 \hline
 Event & At 3-12 m distance & At 40 m distance \\
 \hline
 Rumbles (rms) & 37.49 µV & 22.65 µV \\
 Stomps (Peak to Peak) & 4178.46 µV & 1458.6 µV \\
 \hline
 \end{tabular}
 \label{tab:voltages}
\end{table}

The embedded system utilized \hl{an} ATmega328p  microcontroller, equipped with an ADC featuring a resolution of 4.88 mV. However, the maximum output voltage of the geophone (4178.46 $\mu$V) being lower than the ADC resolution hindered the direct interfacing of a geophone to the embedded system for the detection of useful seismic signal patterns. Similarly, noise interference for seismic signals also made the direct integration of geophones into an embedded system more challenging. Therefore, a combination of a geophone and a microcontroller is insufficient to gather seismic waves from elephants. \hl{Consequently, following basic electronic fundamentals, the geophone-sensor interface depicted in Fig. \mbox{\ref{fig:circuit}} has been introduced between the geophone and the microcontroller (as 
 proposed in Fig. \mbox{\ref{fig:process_diagram})} to address these challenges.}

 \subsection{Geophone-Sensor Interface Design}
 The geophone-sensor interface, as illustrated in Fig. \ref{fig:circuit}, is a cascade of an instrumentation amplifier, a noise filter and a signal amplifier.
 
 \subsubsection{Instrumentation amplifier}
An instrumentation amplifier (IA) was implemented at the beginning of the sensor interface because the amplifier positioned at the beginning of the sensor interface plays a critical role in determining the sensitivity of the geophone-sensor interface (see Fig. \ref{fig:circuit} (a)). The IA is a precision device known for its high input impedance, low output impedance, low self-generated noise, low offset drift, and high Common Mode Rejection Ratio (\hl{CMRR}) \cite{fraser1994electrical}. Importantly, the IA can amplify weak signals significantly without adding significant noise \cite{nagulapalli2019novel}. 

\hl{However, the gain of a general 3-op-amp IA implementation may cause offset drift due to temperature-dependent voltage output \mbox{\cite{fraser1994electrical}}. Also, an unequal alteration of resistance or capacitance ratios within the IA topology adversely affects gain accuracy and CMRR \mbox{\cite{lin2018review}} compromising the performance of the geophone-sensor interface. To mitigate this, a monolithic instrumentation amplifier was used in the proposed system. With this implementation, both active and passive components are on the same die, ensuring close matching and maintaining component matching over temperature changes. Consequently, a high CMRR is attained, and the system is expected to demonstrate excellent performance over a considerable temperature range.}

The AD620 monolithic amplifier was utilized for this study because laser trimming of an on-chip thin film resistor allows for setting the gain to be 100 within $\pm(0.3)\%$ max error \cite{kitchin2006designer}. The gain of the instrumentation amplifier is determined by a single resistor, and it is capable of achieving a wide range of gains from 1 to 10,000. The gain of the instrumentation amplifier $A_{v1}$ can be calculated using equation (\ref{eq:AD620gain}) where $R_G$ is the single resistor that can be used to adjust the gain. \hl{The resistor $R_B$ depicted in Fig. \mbox{\ref{fig:circuit}} (a) is utilized to adjust the amplitude offset of the signal} 

\begin{equation}\label{eq:AD620gain}
 A_{v1\ }=\left(1+\frac{49.4 k\Omega}{R_G}\right) 
\end{equation}

 \subsubsection{Noise Filter}
\hl{The strategy implemented to eliminate the noise component is to filter the noise frequencies and pass only the effective seismic frequency band originating from the locomotion of elephants. It is also anticipated that the application of a low-pass filter within the implementation will serve to minimize the occurrence of signal aliasing.} According to Wood et al., \cite{wood2005using}, the general footfall of animals generates seismic signals below 80 Hz. Seismic waves generated by elephants are expected to be around 20 Hz \cite{o2000seismic}. Seismic noises, such as earthquakes, typically fall within frequencies ranging from 0.01 Hz to 10 Hz \cite{tosi2012earthquake}. Therefore, an ideal filter should pass signals within the frequency range of 10 Hz to 80 Hz to significantly reduce the amount of noise.

\hl{Consequently}, the geophone-sensor interface was designed to pass frequencies around 20 Hz. Therefore, the overall implementation is intended to function as a band-pass filter with a lower cutoff frequency of 10 Hz. 
Although the upper limit of the focused frequency range is 80 Hz, an upper cutoff frequency of 200 Hz was selected making an upper-frequency tolerance \footnote{ During the signal processing stages, digital filters will be employed to adjust the upper cutoff frequency if required.}.

 \begin{figure}[!t]
 \centering
 \includegraphics[width=0.45\textwidth]{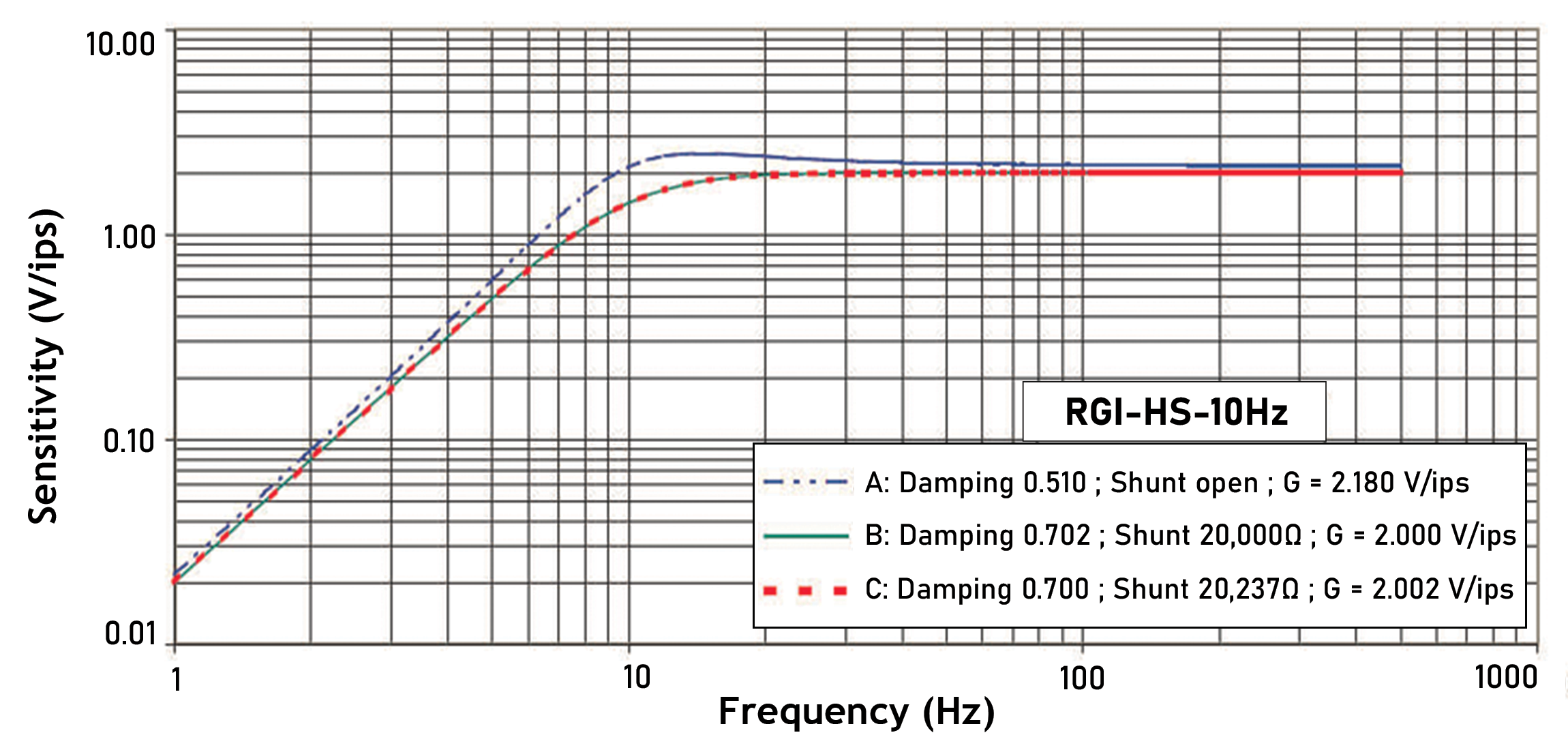}
 \caption{\textbf{Frequency response of geophone (RGI-HS10) \cite{RacotechGeophysicalInstrument}}}
 \label{fig:geophoneresponse}
\end{figure}

As mentioned, the selected geophone (RGI HS 10) exhibits the frequency response as illustrated in Fig. \ref{fig:geophoneresponse}. The natural frequency of the device is 10 Hz. According to Fig. \ref{fig:geophoneresponse}, the geophone's frequency response shows the characteristics of a high-pass filter, and the natural frequency of the geophone serves as the cutoff frequency. In this case, the geophone itself attenuates the signals in \hl{frequencies} below 10 Hz. Therefore, no effort was made in the geophone-sensor interface design to integrate a high-pass filter.

\hl{A second-order Butterworth filter was integrated into the geophone-sensor interface as the low-pass filter. The low-pass Butterworth filter implementation with Sallen-key topology is depicted in Fig. \mbox{\ref{fig:circuit} (b).}} Compared to other filter configurations (such as the Bessel filter or Chebyshev filter), the Butterworth filter exhibits the flattest passband response. Hence, the effect on the signal pattern was minimized. However, it compromises steepness in the transition region from the passband to the stopband \cite{horowitz2015art}. 

 \hl{The Butterworth filter was implemented using the commercially available components such as }$R_1=R_2=R=1\ k\mathrm{\Omega}$, $C_1=C_2=C=1\ \mathrm{\mu}F$ and LM324 Op-Amp. For the quality factor of $ Q= 0.707 $, the commercially available resistor values are approximately selected as $R_3=10\ k\mathrm{\Omega}$ and $R_4=5.1\ k\mathrm{\Omega}$. The cutoff frequency $f_c$ can be calculated using (\ref{eq:cutoffofbw}) where $R$ is the resistor value when $R_1=R_2$ and $ C $ is the capacitor value when $C_1=C_2$. The gain of the filter $A_{v2}$ can be calculated using (\ref{eq:gainofbw}) where $R_4$ and $R_3$ are resistors represented in Fig. \ref{fig:circuit} (b).

\begin{equation}\label{eq:cutoffofbw}
 f_c=\frac{1}{2\pi R C}
\end{equation}

\begin{equation}\label{eq:gainofbw}
 A_{v2}=\ 1+\frac{R_4}{R_3}
\end{equation}

Based on the selected values, the cutoff frequency was obtained as $f_c = 159.15\ Hz$ and the gain $A_{v2} = 1.51$. When the filter was simulated, -3 dB magnitude was achieved at $212.064\ Hz$, and the error between the targeted cutoff frequency and obtained cutoff frequency was neglected as the targeted cutoff frequency was loosely defined with a tolerance. 

\subsubsection{Signal Amplifier}
The signal amplifier, the final stage of the geophone-sensor interface, is responsible for amplifying the signal to optimized voltage levels. In the design, as illustrated in Fig. \ref{fig:circuit} (c), a non-inverting amplifier with a fixed gain was implemented using an LM324 Op-Amp, and the gain $A_{v3}$ was defined according to (\ref{eq:signalgain}).

\begin{equation}\label{eq:signalgain}
 A_{v3}=1+\frac{R_f}{R_{in}}
\end{equation}

The field-tested prototype \hl{was designed} with $R_f =10\ k\mathrm{\Omega}$ and $R_{in} =1\ k\mathrm{\Omega}$. Hence the signal amplifier has a fixed gain of 11.

Finally, the overall gain of the geophone-sensor interface is $A_{v(total)}$, calculated using (\ref{eq:overallgain}), where $A_{v1}$ is the gain of the instrumentation amplifier, $A_{v2}$ is the gain of the Butterworth filter, and $A_{v3}$ is the gain of the signal amplifier. Given the fixed gains used in the filter and signal amplifier design, $A_{v2} \times A_{v3}$ can be calculated as $16.61$. As $A_{v1}$ can vary from 1 to 10,000, $A_{v(total)}$ can be theoretically calculated between 16.61 to 166,100 using (\ref{eq:overallgain}). The seamless integration of a geophone-sensor interface into the instrumentation system enables a steady passband response of seismic signals within the band of 10 Hz to 212.064 Hz.

\begin{equation}\label{eq:overallgain}
 A_{v(total)}\ =A_{v1} \times A_{v2} \times A_{v3}
\end{equation}

\subsection{Prototype Implementation}

\begin{figure}[!t]
 \centering
 \includegraphics[width=0.4\textwidth]{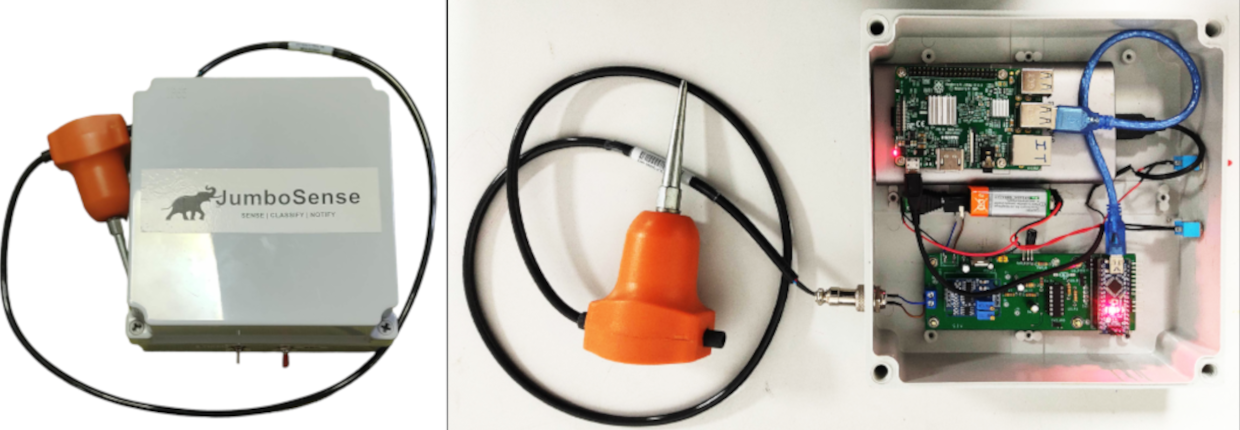}
 \caption{\hl{\textbf{Integrated prototype}} }
 \label{fig:prototype}
\end{figure}

\begin{figure}[!t]
 \centering
 \includegraphics[width=0.22\textwidth]{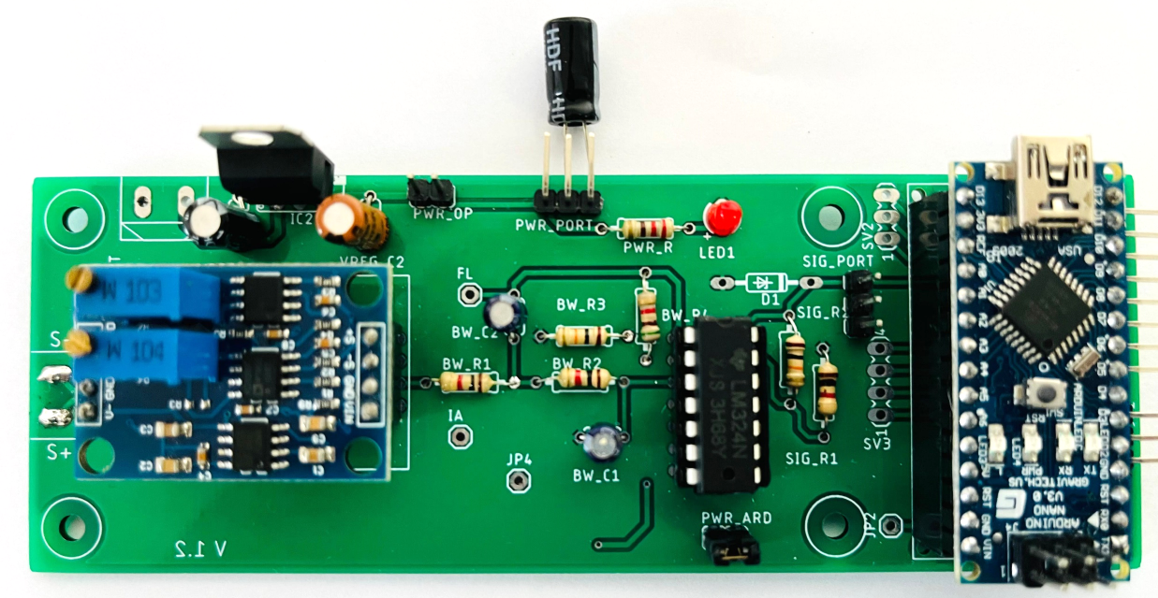}
 \caption{\hl{\textbf{Geophone-sensor interface}} }
 \label{fig:GSI}
\end{figure}

\hl{The physical prototype of the geophone-sensor interface implementation is shown in Fig. \mbox{\ref{fig:prototype}} and the latest version of the geophone-sensor interface is shown in Fig. \mbox{\ref{fig:GSI}}.} 
The geophone is connected to the geophone-sensor interface using a 1 m long cable, whereas the rest of the components: the geophone-sensor interface, the microcontroller, and a battery system were compacted as a single unit. The components were enclosed within a protective enclosure to make them suitable for outdoor use. The microcontroller has an extendable connection for the laptop which \hl{was} used for data recording and visualization \footnote{\hl{For unattended data collection during field experiments, the laptop can be replaced with a Raspberry Pi device.}}.

\hl{The prototype was implemented using an Arduino development board with an AVR ATmega328p microcontroller IC. ATmega328p  is an 8-bit, high-performance, low-power microcontroller.} The output signal of the geophone-sensor interface was connected to the internal ADC, a 10-bit ADC with a resolution of 4.88 mV. The microcontroller \hl{was operated} with a 16 MHz \hl{oscillator}, and the default 128 prescaler was used, making the clock speed 125 kHz. Given that the microcontroller requires 13 clock cycles for ADC conversion, the maximum frequency for ADC can operate with the given prescaler is 9,615 Hz.

When implementing the prototype, serious attention \hl{was paid} to the effect created by changing temperatures, given that the proposed method will be planted and tested in a tropical outdoor environment in Sri Lanka. Therefore, the average yearly temperature was considered to be in the range between 26°C to 28°C, and the day and night temperatures were considered to vary by 4°C to 7°C from the average temperature \mbox{\cite{de2006impacts}}. 

The development board was programmed using C++, and during the programming stage, several delays had to be introduced to the program to ensure smooth data acquisition and transmission processes. Considering introduced delays and the slight variations in \hl{the} program execution time, the approximated average sampling frequency was limited to 880 Hz. In the data recording setup, the development board was connected to a laptop, and a multithreaded Python script was employed to collect serial data, \hl{to save} it as a CSV file, and \hl{to enable} real-time plotting of the seismic signals.

\section{Simulations, Field Experiments, Results and Discussion}

A series of simulations and field experiments were carried out to evaluate the geophone-sensor interface's stability and \hl{to determine} its accuracy and capacity to detect elephant locomotion at long ranges. 
\begin{itemize}
    \item Initially, the designed electronic circuit was assessed for its performance and stability using simulation software: NI Multisim 12.0;
    \item \hl{Secondly, the completed geophone-sensor interface was tested for applicability using human subjects;}
    \item \hl{Finally, it was used to read the seismic signal pattern of the elephant's locomotion and to identify the detection range.}
\end{itemize}

All the experiments \hl{that} involved humans were conducted after explaining the full process and with their consent. Similarly, all experiments involving elephants were conducted under the strict supervision of the Department of National Zoological Gardens in Sri Lanka. The examination of equipment and experimental methodologies was carried out to ensure the absence of any significant adverse effects on the elephants. Additionally, the participation of the elephants in the study was facilitated by designated caretakers and mahouts assigned by the Department of National Zoological Gardens.

\subsection{Software Simulation}

The preliminary software simulations were carried out to verify the capability of the geophone-sensor interface to detect elephants. \hl{Asian elephants' footfalls from a 40 m distance presented in \mbox{\cite{o2000seismic}} were assumed as the baseline scenario for the simulation}. The geophone should generate 1458.6 µV peak-to-peak voltage for this case, according to the data presented in Table \ref{tab:voltages}. However, as the actual locomotion signal pattern is impossible to simulate using software, \hl{a sine signal of amplitude = 1458.6 µV and frequency= 20 Hz was used for simulation.} 

The input signal was amplified using (\ref{eq:gainforADC}), where $A_{v(total)}$ is the required gain for the geophone-sensor interface, $V_{ADC}$ is the input voltage range of the ADC, and $V_{Signal}$ is the peak-to-peak voltage of the input signal.\hl{For the selected scenario, the input voltage range ($V_{ADC}$) of ATmega328p is 5 V, the signal voltage ($V_{Signal}$) is 1458.6 µV and the calculated $A_{v(total)}$ is 3427.9.} 

\begin{equation}\label{eq:gainforADC}
 A_{v(total)} =\frac{V_{ADC}}{V_{Signal}}
\end{equation}

\hl{However, the output voltage swing of LM324 is $0 V$ to $VCC- 1.5 V$. This limits the maximum voltage range of the geophone-sensor interface from  0 to 3.5 V. Therefore, with the calculated gain of 3,427.9, the signals above 3.5 V were clipped.} To overcome this issue, the simulation was conducted for an approximated gain ($A_{v(total)}$) of 2,000, which is less than the theoretically calculated gain. Consequently, using  (\mbox{\ref{eq:overallgain}}), the ($A_{v1}$) was calculated as 120.41 for the simulation purposes.

 \begin{figure}[!t]
 \centering
 \includegraphics[width=0.38\textwidth]{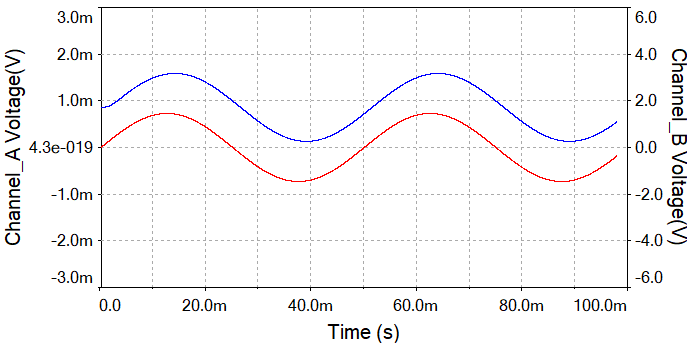}
 \caption{\textbf{Simulated time domain response of geophone-sensor interface for 20 Hz sine signal (channel A: red, channel B: blue)}}
 \label{fig:simulatedoutputosci}
\end{figure}

The simulation results are presented in Fig. \ref{fig:simulatedoutputosci}; channel A (red) represents the input sine signal and channel B (blue) is the output signal. As shown in the figure, the geophone-sensor interface can amplify the simulated signal to an acceptable voltage range (to 2.909 V peak-to-peak signal). The signal pattern was recovered with no errors. However, a slight phase difference was observed and that is not abnormal for a geophone-sensor interface. 

\begin{figure}[!t]
 \centering
 \includegraphics[width=0.45\textwidth]{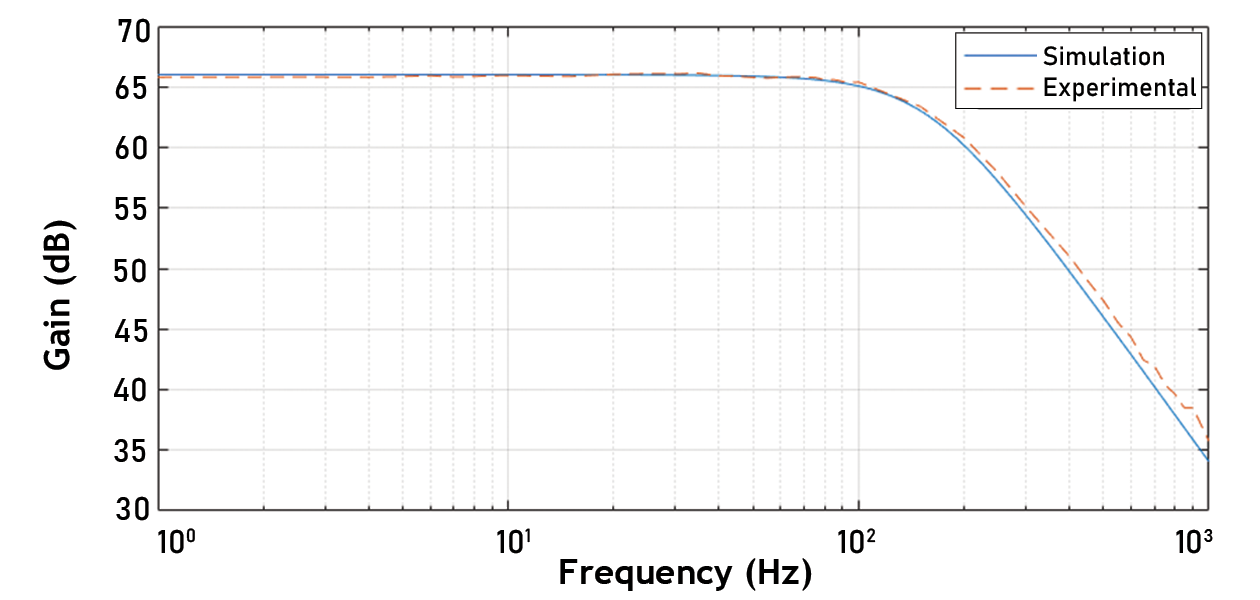}
 \begin{center}
 (a)
 \end{center}
 \includegraphics[width=0.45\textwidth]{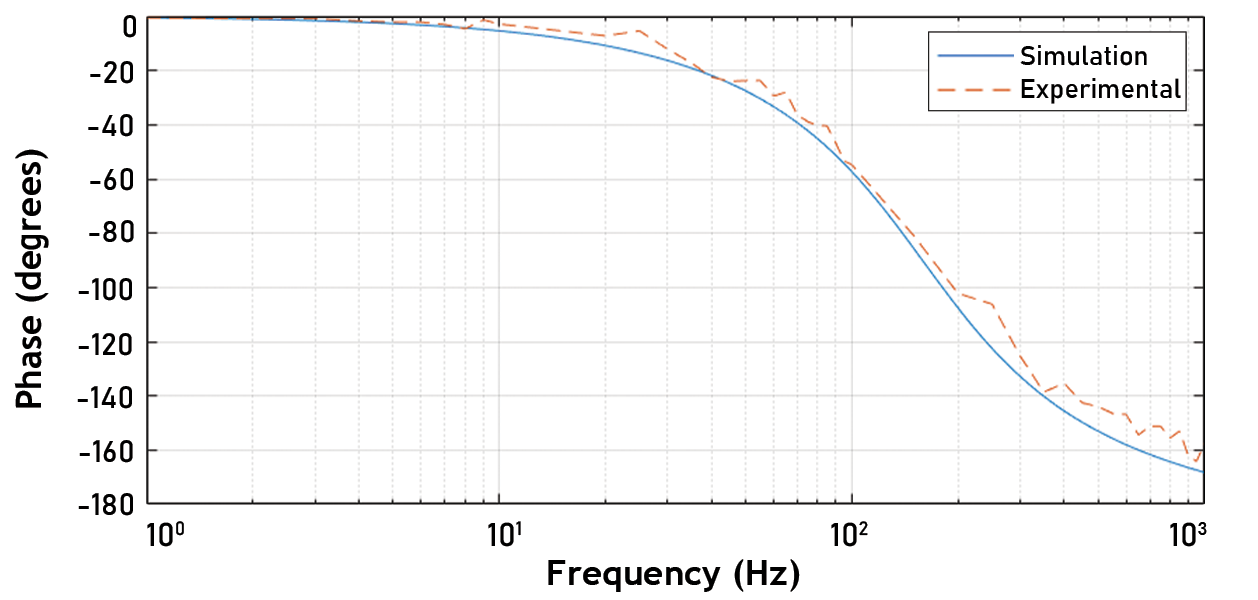}
 \begin{center}
 (b)
 \end{center}
 \caption{\hl{\textbf{(a) Gain of the geophone-sensor interface ($A_{v(total)}$) for the frequency range 1 Hz to 1 kHz (b) Phase shift of the system for the frequency range 1 Hz to 1 kHz}}}
 \label{fig:bodeplot}
\end{figure}

To evaluate the stability of the geophone-sensor interface, the open loop magnitude response and phase response were simulated using a bode plot. The same gain configurations were assumed for this simulation, and the frequency response was analyzed for the frequency range from 1 Hz – 1 kHz. \hl{ The results are presented in Fig. \mbox{\ref{fig:bodeplot}} and according to the figure, in general, it can be observed that the simulation and experimental values are approximately matched, confirming the proper functioning of the system.}

Further, \hl{the} analysis of Fig. \mbox{\ref{fig:bodeplot}} (a) shows \hl{that} the gain is an approximately flat response for the frequency range between 10 Hz to 80 Hz, which is the frequency range considered in this study. The average gradient for the gain in the observed 70 Hz band is $-6.9622 \times 10^{-3}$. When considering Fig. \mbox{\ref{fig:bodeplot}} (b), a slight variation is observed within the considered frequency range, and it is approximately 39.8310 degrees. Yet,  the overall gain difference and the phase shift demonstrate acceptable differences, signifying \hl{the} geophone-sensor interface is stable in detecting seismic signals of elephants.

\begin{figure}[!t]
 \centering
 \includegraphics[width=0.42\textwidth]{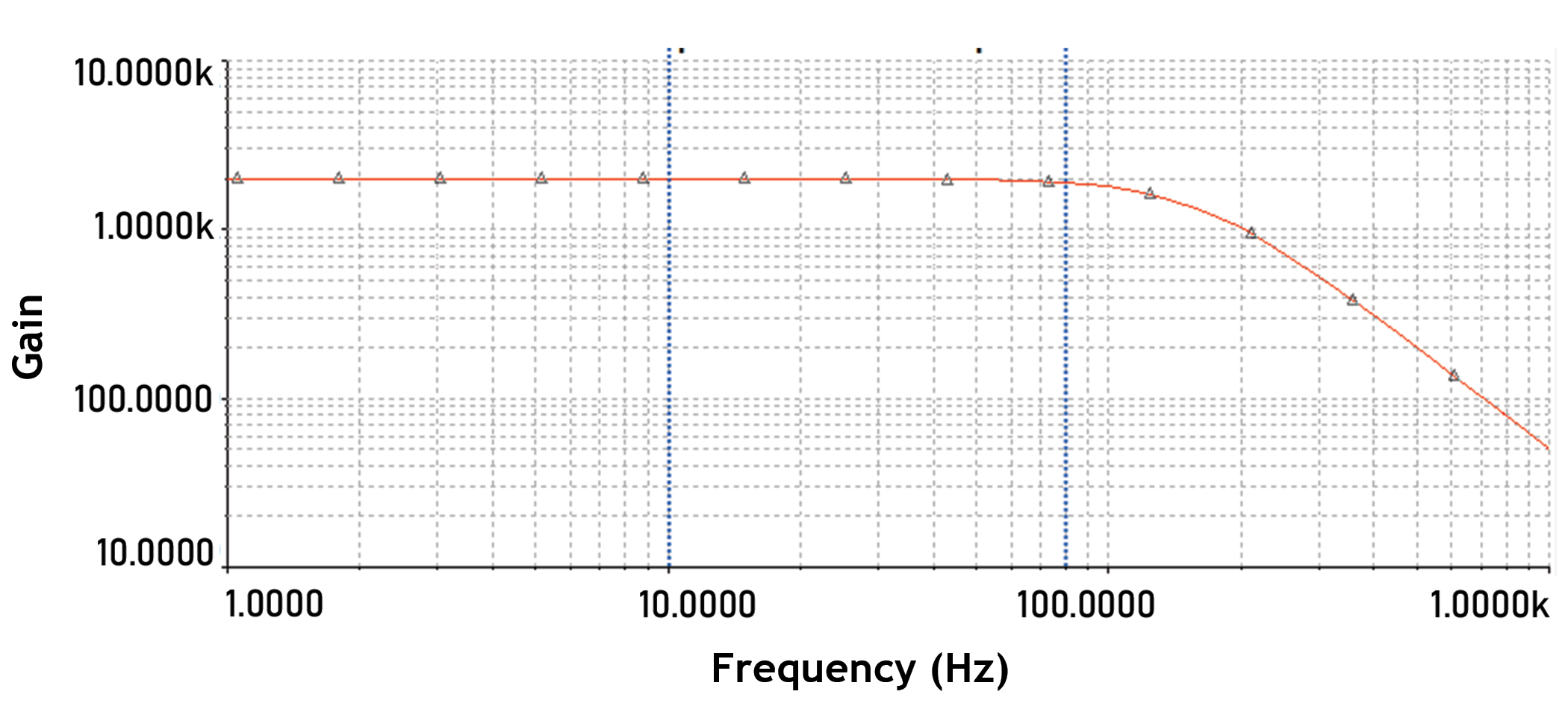}
 \begin{center}
 (a)
 \end{center}
 \includegraphics[width=0.42\textwidth]{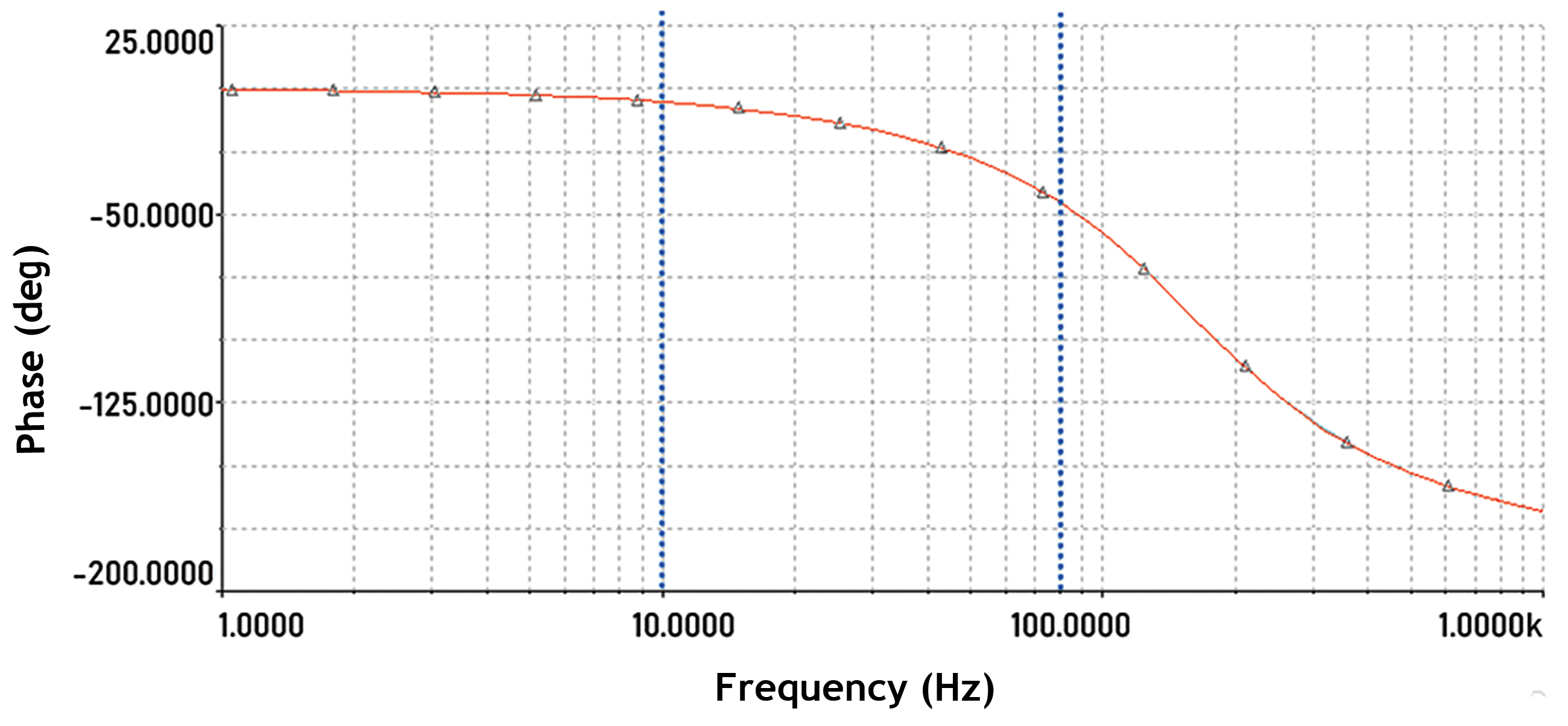}
 \begin{center}
 (b)
 \end{center}
 \caption{\hl{\textbf{(a) Variation of gain of the geophone-sensor interface across the frequency range 1 Hz to 1 kHz at temperatures ranging from 10°C to 40°C (b) Variation of phase shift of the geophone-sensor interface across the frequency range 1 Hz to 1 kHz at temperatures ranging from 10°C to 40°C.} }}
 \label{fig:tempanalysis}
\end{figure}

The design of the geophone-sensor interface was tested for temperature variation ranging from 10°C to 40°C in 1°C increments, along with AC analysis covering the frequency range from 1 Hz to 1 KHz. The yielded results are depicted in Fig. \ref{fig:tempanalysis}. As presented in \hl{the} Figure, it is important to note that the simulated results for different temperatures could not be distinguished separately because all 31 temperature variations \hl{are overlapping}. Consequently, the results demonstrate that the circuit remains stable within the temperature range of 10°C to 40°C. These simulation results assure that the introduced geophone-sensor interface can operate stably and accurately as expected in average day-to-night temperature variations in targeted tropical countries, i.e., Sri Lanka.

\subsection{Field Experiments}
The field experiments were carried out in four stages (A-D) to identify human footfalls, elephant footfalls, and maximum detection range for elephants respectively. In particular, Experiment D was carried out in the area that has the most significant instances of HEC \cite{prakash2020human}. All field experiments were conducted assuming the soil conditions as given in \cite{Agriculture_Maps_2023}\footnote{The legend of the original figure in \mbox{\cite{Agriculture_Maps_2023}} has been improved to enhance the readability of the image.}. During the experiments temperature variances from 24°C (Experiment C) to 31°C (Experiment A), and humidity variances from 64\% (Experiment A) to 84\% (Experiment C) were observed.

\begin{figure}[!t]
 \centering
 \includegraphics[width=0.4\textwidth]{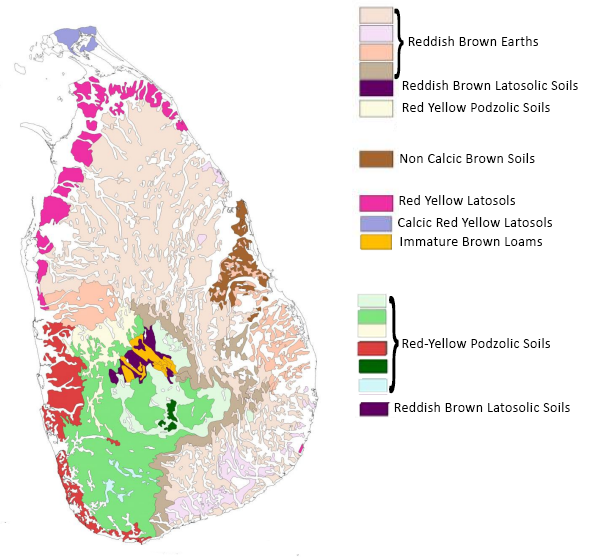}
 \caption{\textbf{Soil map of Sri Lanka\mbox{\cite{Agriculture_Maps_2023}}}}
 \label{fig:Soil Map of Sri Lanka}
\end{figure}

\subsubsection{Experiment A}
\begin{figure}[!t]
 \centering
 \includegraphics[width=0.4\textwidth]{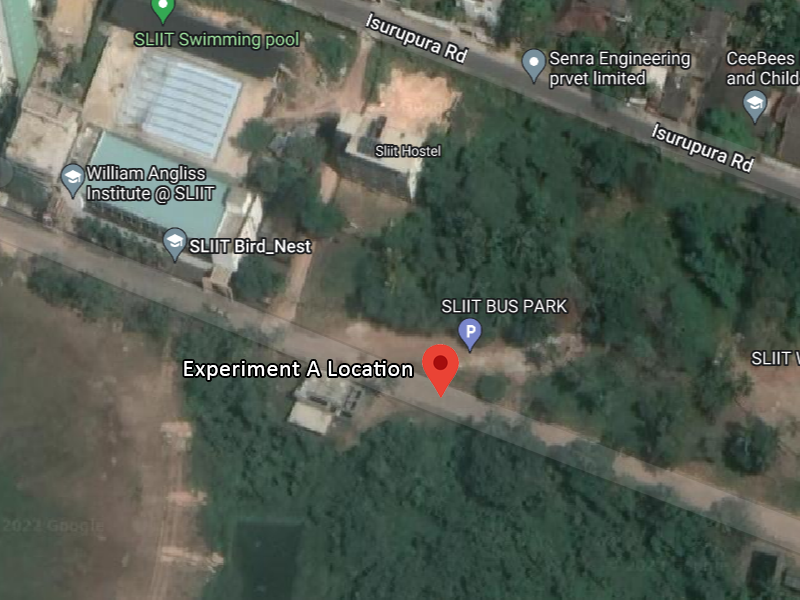}
 \caption{\textbf{Site of experiment A - Sri Lanka Institute of Information Technology outdoor premises } }
 \label{fig:locA}
\end{figure}

The first field experiment (\hl{hereinafter referred to} as Experiment A)  was conducted in two stages aiming to verify the system's behavior in response to general ambient activities and to detect seismic signals related to human behavior. The details of the selected human subjects are provided in Table \ref{tab:subjectexpA}. In addition to human subjects, a motorcycle was also included as an additional research subject. For both experiments, sampling rates were set as 8,835 Hz and the overall gain of the sensor interface ($A_{v(total)}$) was set to 580. The experiments were conducted in the outdoor environment of the lab premises at the Sri Lanka Institute of Information Technology in Malabe, Sri Lanka (see Fig. \ref{fig:locA}).

\begin{figure*}[t]
 \centering
 \begin{minipage}[t]{0.4\textwidth}
   \centering
   \includegraphics[width=\textwidth]{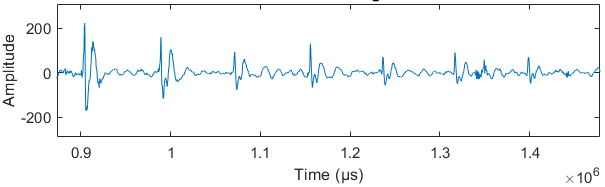}
   \centering (a)
 \end{minipage}
 \hfill
 \begin{minipage}[t]{0.4\textwidth}
   \centering
   \includegraphics[width=\textwidth]{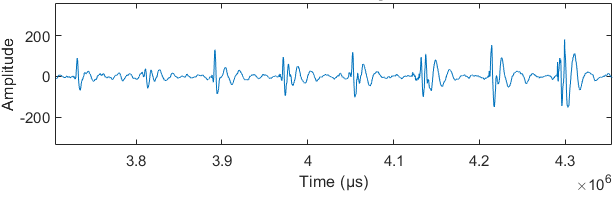}
   \centering (b)
 \end{minipage}
 \vskip\baselineskip
 \begin{minipage}[t]{0.4\textwidth}
   \centering
   \includegraphics[width=\textwidth]{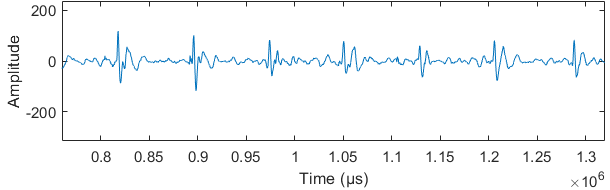}
   \centering (c)
 \end{minipage}
 \hfill
 \begin{minipage}[t]{0.4\textwidth}
   \centering
   \includegraphics[width=\textwidth]{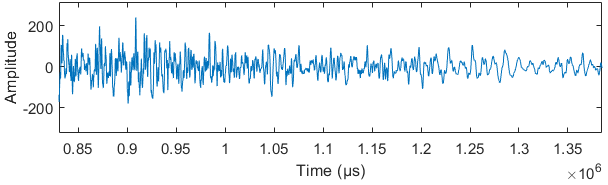}
   \centering (d)
 \end{minipage}
\caption{\textbf{Time domain representation of (a) Walking pattern of the subject P3, (b) Walking pattern of the subject P4, (c) Walking pattern of the subject P5, (d) Motorcycle} }
 \label{fig:anyseimicsignal}
\end{figure*}

In the first stage of Experiment A, the system's detection range was assessed by simultaneous jumps of P1, P2, P4, and P5. Although the maximum reported distance was 48.5 m, the simultaneous jumps demonstrated a saturated signal; in most attempts, the signals were clipped, resulting in no observable signal pattern. For the second stage of Experiment A, P3, P4, and P5 were asked to walk in a straight line for 15 m. The seismic signal patterns for each person were recorded individually. The seismic signal patterns for the footfalls of P3, P4, and P5 are depicted in Fig. \ref{fig:anyseimicsignal} (a), Fig. \ref{fig:anyseimicsignal} (b), and Fig. \ref{fig:anyseimicsignal} (c), respectively. In addition to the walking patterns, the signal pattern of a motorcycle was also recorded in Fig. \ref{fig:anyseimicsignal} (d)\footnote{\hl{The figures depict amplitudes corresponding to ADC readings; as such, specific units were not provided.}}.

A visible time domain pattern for human footfalls can be observed in Fig. \ref{fig:anyseimicsignal}, \hl{and it confirms} that the geophone-sensor interface successfully acquires seismic signal patterns corresponding to human gaits. It is also noted that the reported predominant frequency for human walking ranges from 60.63 Hz to 70.68 Hz. However, a tendency to record false values was observed in some instances. Therefore, as mentioned in the methodology, to address this issue, the sampling frequency was reduced from 8,835 Hz to approximately 880 Hz in the later experiments. According to Fig. \ref{fig:anyseimicsignal} (d), unlike the human walking signals, no visible pattern can be identified and it is considered an observation of noise. The predominant frequency observed was 71.38 Hz. 
\begin{table}[t]
 \centering
 \caption{\textbf{Human test subjects used for experiment A}}
 \begin{tabular}{|c|c|c|}
 \hline
 Test Subject & Gender & Weight (Kg) \\
 \hline
 P1 & Female & 67 \\
 P2 & Female & 65 \\
 P3 & Female & 60 \\
 P4 & Male & 90 \\
 P5 & Male & 78 \\
 \hline
 \end{tabular}
 \label{tab:subjectexpA}
\end{table}

\subsubsection{Experiment B}
\begin{figure}[!t]
 \centering
 \includegraphics[width=0.4\textwidth]{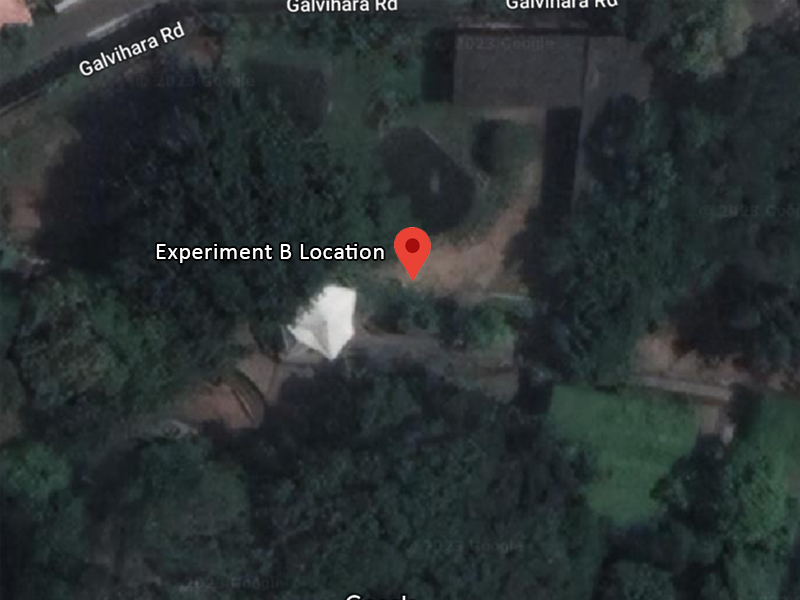}
 \caption{\textbf{Site of experiment B - National Zoological Gardens, 
Dehiwala, Sri Lanka} }
 \label{fig:locB}
\end{figure}

The second experiment (\hl{hereinafter} referred to as Experiment B) was conducted using three tamed elephants from the National Zoological Gardens, Dehiwala, Sri Lanka (see Fig. \ref{fig:locB}), to verify the behavior of the geophone-sensor interface for elephants in a completely controlled environment. The physical characteristics of the selected elephants are presented in Table \ref{tab:subjectexpB}. Some data are missing due to physical limitations; for example, a special scale should be used to measure elephant weight, which was impossible given the setup of the zoological garden.

\begin{table}[!t]
 \centering
 \caption{\textbf{Physical properties of elephants used in experiment B}}
 \begin{tabular}{|p{1cm}|p{1cm}|p{1cm}|p{1cm}|p{2.2cm}|}
 \hline
 Elephant & Age (Years) & Weight (Kg) & Height (cm) & {Front Leg Circumference (cm)} \\
 \hline
 D1 & 22 & - & 198 & 111 \\
 D2 & 40 & - & 236 & 127 \\
 D3 & 45 & - & - & 122 \\
 \hline
 \end{tabular}
 \label{tab:subjectexpB}
\end{table}

\begin{figure}[!t]
 \centering
 \includegraphics[width=0.38\textwidth]{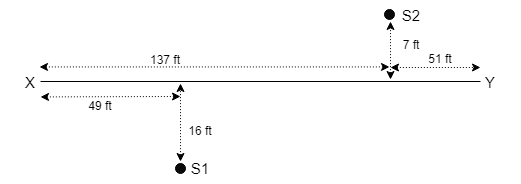}
 \caption{\textbf{Sensor placement for experiment B} }
 \label{fig:dehiwalasetup}
\end{figure}

\begin{figure}[!t]
 \centering
 \includegraphics[width=0.4\textwidth]{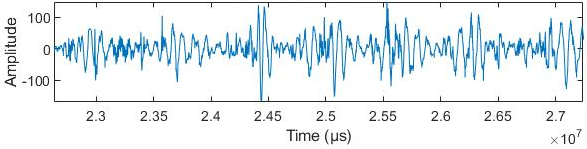}
 \caption{\textbf{Time domain signal representation of elephant walking in experiment B} }
 \label{fig:dehiwalawalking}
\end{figure}

As illustrated in Fig. \ref{fig:dehiwalasetup}, the area provided for Experiment B was 64.3 m long and considerably noisy due to \hl{the presence of} visitors to the zoo. The sensors were buried (nearly 7 cm \hl{deep}) in two different places (S1 or S2 in Fig. \ref{fig:dehiwalasetup}) to conduct two rounds of experiments considering both sides of the test area. The elephants were made to walk and run in an approximately straight line. The same gain used in Experiment A ($A_{v(total)}=580$) was used for this experiment as well. The observed results are presented in  Fig. \ref{fig:dehiwalawalking}.

The geophone-sensor interface provided a substantial response to the walking signals of elephants, as visually observable in Fig. \ref{fig:dehiwalawalking}. However, due to the location being in an urban area with high noise levels, the signal patterns were highly affected by the noise. It was also observed that the space for the experiments was insufficient (distance between S1 and S2 in Fig. \ref{fig:dehiwalasetup}). \hl{Despite these limitations}, the primary objective of this \hl{experiment, i.e.,}testing the geophone-sensor interface for actual elephants to gather seismic waves, was successfully achieved.

\subsubsection{Experiment C}
\begin{figure}[!t]
 \centering
 \includegraphics[width=0.5\textwidth]{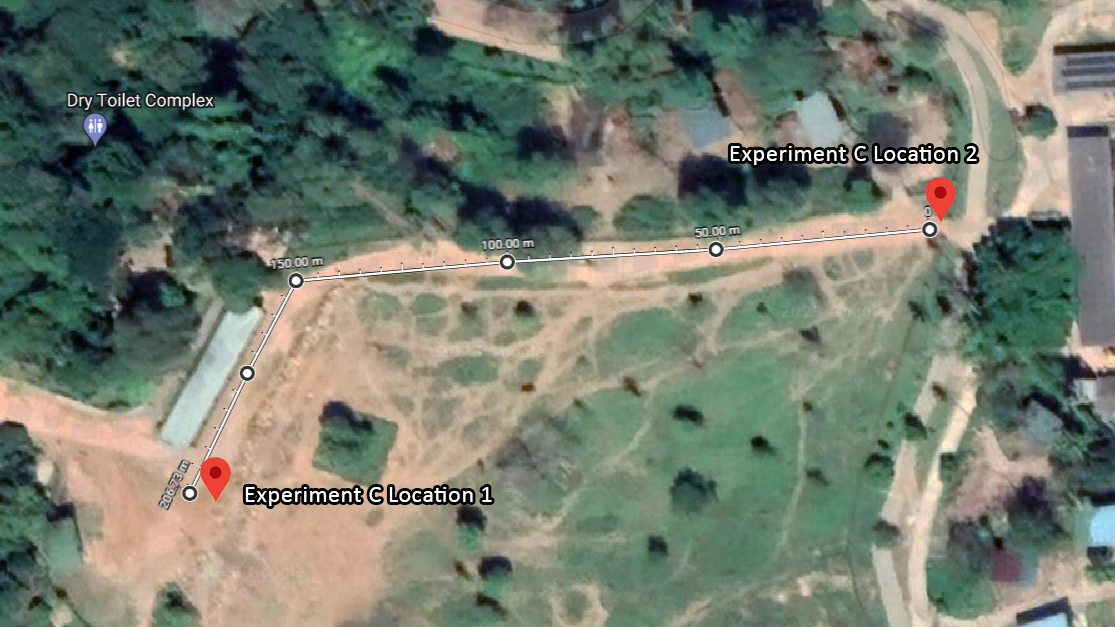}
 \caption{\textbf{Site of experiment C - Pinnawala elephant orphanage in Rambukkana, Sri Lanka} }
 \label{fig:locC}
\end{figure}

The third experiment (\hl{hereinafter} referred to as Experiment C) was conducted in the Pinnawala Elephant Orphanage in Rambukkana, Sri Lanka (see Fig. \ref{fig:locC}), to avoid the main problems encountered in Experiment B: the noisy environment and insufficient distance. The primary objectives of Experiment C were to study the maximum detection range of elephants' footfalls and to benchmark the system, specifically the gain, in acquiring the seismic signal patterns from the elephants' footfalls within an environment close to their natural habitats. Three partly tamed elephants were provided for this experiment, and their physical characteristics are presented in Table \ref{tab:subjectexpC}. 

\begin{table}[!t]
 \centering
 \caption{\textbf{Physical properties of elephants used as test subjects (experiment C)}}
 \begin{tabular}{|p{1cm}|p{1cm}|p{1cm}|p{1cm}|p{2.2cm}|}
 \hline
 Elephant & Age (Years) & Weight (Kg) & Height (cm) & {Front Leg Circumference (cm)} \\
 \hline
 E1 & 56 & 3500 & 233 & 129 \\
 E2 & 25 & 3100 & 237 & 127 \\
 E3 & 16 & 2700 & 222 & 110 \\
 \hline
 \end{tabular}
 \label{tab:subjectexpC}
\end{table}

\begin{figure*}[!t]
 \centering
 \begin{minipage}[t]{0.3\textwidth}
   \centering
   \includegraphics[width=\textwidth]{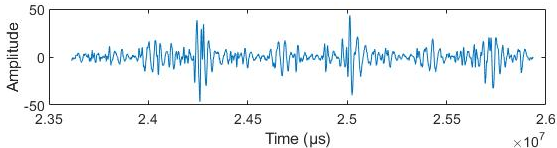}
   (a)
 \end{minipage}
 \hfill
 \begin{minipage}[t]{0.3\textwidth}
   \centering
   \includegraphics[width=\textwidth]{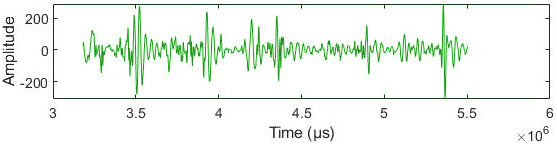}
   (b)
 \end{minipage}
 \hfill
 \begin{minipage}[t]{0.3\textwidth}
   \centering
   \includegraphics[width=\textwidth]{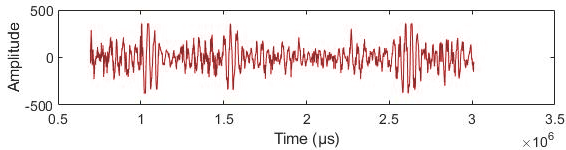}
   (c)
 \end{minipage}
 \caption{\textbf{Walking pattern of elephant Under (a) gain S0 - blue,  (b) gain S1 - green, and (c) gain S2 - red}}
 \label{fig:comparegaintime}
\end{figure*}

\begin{figure*}[!t]
 \centering
 \minipage{0.28\textwidth}
 \includegraphics[width=0.8\textwidth]{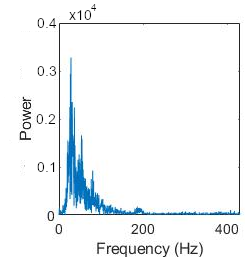}
 \begin{center}
 (a)
 \end{center}
 \endminipage\hfill
 \minipage{0.28\textwidth}
 \includegraphics[width=0.8\textwidth]{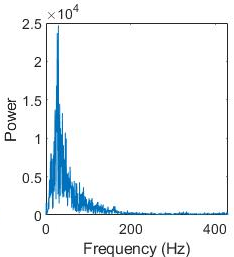}
 \begin{center}
 (b)
 \end{center}
 \endminipage\hfill
 \minipage{0.28\textwidth}
 \includegraphics[width=0.75\textwidth]{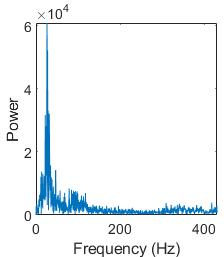}
 \begin{center}
 (c)
 \end{center}
 \endminipage\hfill
 \caption{\textbf{ Power spectrum for walking pattern of elephant under (a) gain S0, (b) gain S1, and (c) gain S2} }
 \label{fig:comparegainfreq}
\end{figure*}

\begin{figure}[!t]
 \centering
 \includegraphics[width=0.5\textwidth]{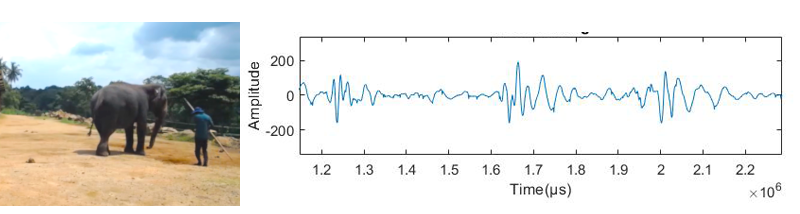}
 \caption{\textbf{Walking pattern of elephant E1 synced with video (location 1, gain setting S1) } }
 \label{fig:walksynked}
\end{figure}

As depicted in Fig. \ref{fig:locC}, two locations were chosen for the experiments: location 1 with a 100 m straight path and location 2 with a 160 m straight path. The sensor was positioned at the corners of the track indicated by the location marks in Fig. \ref{fig:locC}. The experiments were conducted in a partially controlled environment because the three elephants were allowed to behave relatively freely. In all experiments, as illustrated in Fig. \ref{fig:walksynked}, video footage was also utilized to visually synchronize the pattern of the elephants' seismic signals, which is considered the ground truth of the experiments.

Three experiments were carried out at location 2 to measure the maximum detection range. As given in Table \ref{tab:responsetosettings}, the experiments were conducted using three gain settings ($A_{v(total)}$) using elephant E3. The elephant \hl{was made to walk} up to 160 m, and the best observed time domain responses of the geophone-sensor interface for each gain setting are shown in Figure \ref{fig:comparegaintime}.

\begin{table}[!t]
 \centering
 \caption{\textbf{Response of the geophon-sensor interface to different gain settings and distances}}
 \begin{tabular}{|p{1.5cm}|p{1.2cm}|p{1.2cm}|p{1.2cm}|p{1.2cm}|}
 \hline
 Test \newline Configuration & Gain ($A_{v(total)}$) & 0m - 50m & 50m - 100m & 100m - 160m \\
 \hline
 S0 & 580 & Detected & Not Detected & Not detected \\
 \hline
 S1 & 940 & Detected & Detected & Not detected \\
 \hline
 S2 & 25,911 & Saturated & Saturated & Detected \\
 
 \hline
 \end{tabular}
 \label{tab:responsetosettings}
\end{table}

Further to that, the power spectrum representations are shown in Fig. \ref{fig:comparegainfreq}. When considering the power spectrums, the reported peaks are 28.4 Hz, 29.3 Hz, and 26.28 Hz for gain setting S0, gain setting S1 and gain setting S2 respectively. The power of each case showed a difference according to the gain setting and the distance of the elephant from the sensor. Yet, the similarities of the signal patterns and frequencies at peak values in the power spectrum \hl{indicate} that the instrumentation amplifier is capable of detecting seismic signals in all three gain settings without a major distortion of signal patterns.

The maximum distance was reported when the gain setting was S2: $A_{v(total)}=25,911$, and the reported maximum distance was 155.6 m. \hl{The detection range was limited only by the available physical space in the experiment site, and much higher gains would have led to further extension of the distance.} A close observation of Fig. \ref{fig:comparegaintime} (c) incurs a slight clipping at the peak point revealing the elephant was too close to the sensor, even at the maximum tested distance. This observation also confirms that the system's detection range can be further extended. However, as shown in Fig \ref{fig:comparegaintime} (c), the system amplified noise when using higher gain settings. This is because, although the path was isolated at the experiment site, there were other elephants and some human activities observed within a 200 m radius, which may have affected as noise sources for the system. 

\begin{figure}[!t]
 \centering
 \includegraphics[width=0.40\textwidth]{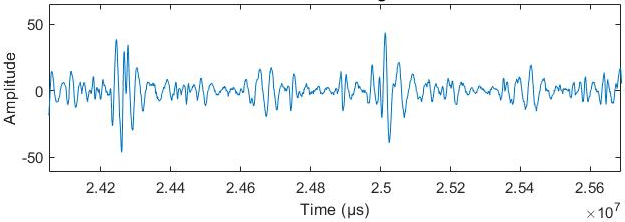}
 \caption{\textbf{Walking pattern of elephant E3 (location 2, gain setting S0 ) } }
 \label{fig:bestpattern}
\end{figure}

\begin{figure}[!t]
 \centering
 \includegraphics[width=0.3\textwidth]{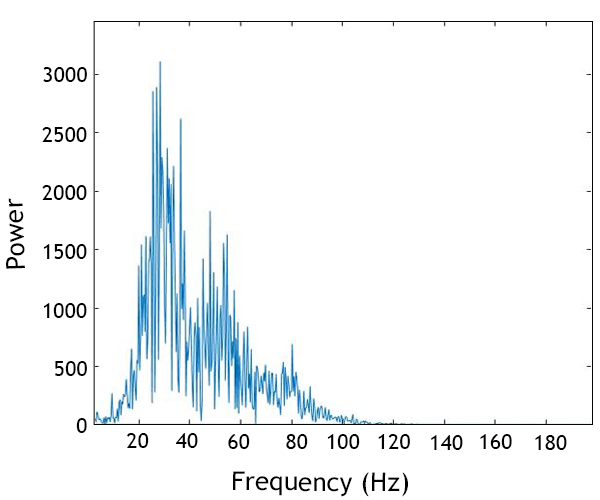}
 \caption{\textbf{Power spectrum of locomotion of elephant (location 2, gain setting S0) } }
 \label{fig:bestpatternspectrum}
\end{figure}

Elephants E1, E2, and E3 were used to carry out several other experiments at location 2 to benchmark the geophone-sensor interface when acquiring seismic signal patterns to identify elephant behaviour. The lowest gain setting S0 was used. Upon comparing the seismic signal patterns, it was observed that the best patterns were exhibited by elephant E3 because the mahout was riding on top of the elephant. \hl{As one of the findings of this study, the optimal observation of the walking pattern of Sri Lankan elephant is presented in Fig. \mbox{\ref{fig:bestpattern}} \footnote{\hl{As per the best knowledge of the authors this is the first time such signal pattern is presented.}} and the corresponding power spectrum is shown in Fig. \mbox{\ref{fig:bestpatternspectrum}}.}

\subsubsection{Experiment D}
\begin{figure}[!t]
 \centering
 \includegraphics[width=0.4\textwidth]{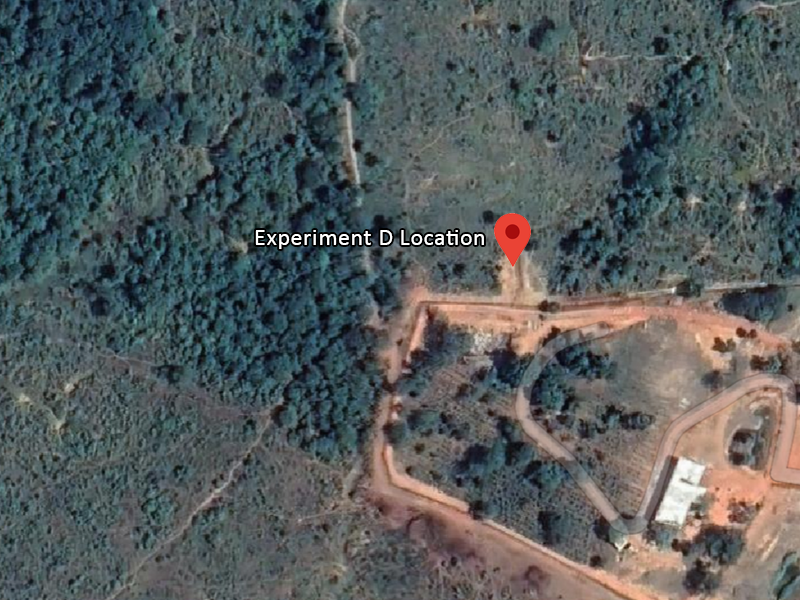}
 \caption{\textbf{Site of experiment D-Digampathaha, Sri Lanka (\hl{geographic coordinates: 7°57'44.8"N 80°41'59.7"E) }}}
 \label{fig:LocD}
\end{figure}

The geophone-sensor interface was used with untamed elephants in their natural habitat as the fourth experiment (hereinafter referred to as Experiment D). This experiment was conducted at the edge of a forest border area in Digampathaha (see \mbox{Fig. \ref{fig:LocD}}), situated within the Central Province of Sri Lanka, adjacent to the borders of the North-Central Province.

\begin{figure}[!t]
 \centering
 \includegraphics[width=0.4\textwidth]{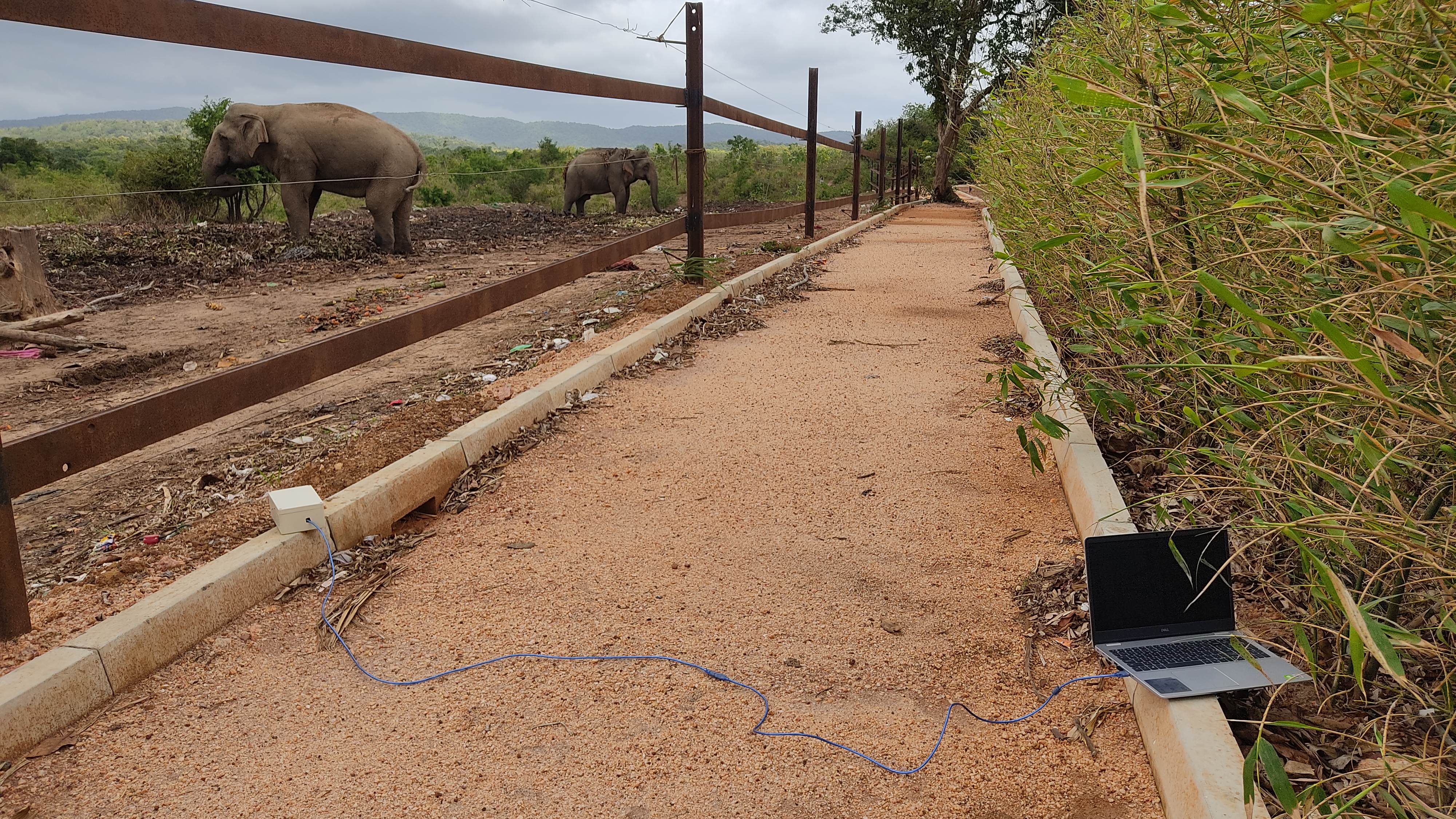}
 \caption{\textbf{Experiment D setup } }
 \label{fig:Dsetup}
\end{figure}

\begin{figure}[!t]
 \centering
 \includegraphics[width=0.45\textwidth]{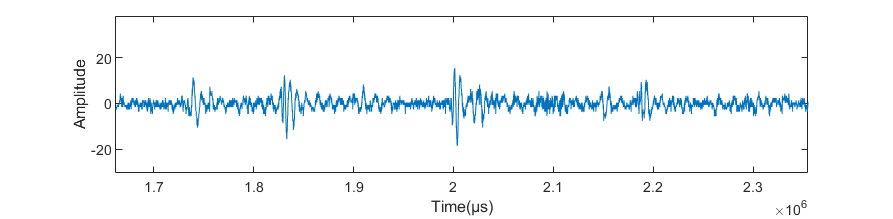}
 \caption{\textbf{Time domain signal representation of elephant walking in experiment D } }
 \label{fig:DPattern}
\end{figure}

As depicted in Fig. \ref{fig:Dsetup}, the geophone-sensor interface was buried adjacent to an electric fence to observe seismic signals of the natural behavior of elephants. The gain ($A_v(total)$) was set to a minimum of 580 because the elephants were near the geophone-sensor interface. The captured locomotion signal patterns of minimal behavior are shown in Fig. \ref{fig:DPattern}. According to the figure, the signal pattern exhibits a significantly low amplitude recurring pattern, confirming that the geophone-sensor interface is capable of successfully observing the minimal locomotive activities of elephants.

\subsection{Accuracy in Elephant Detection}
\label{subsec:Accuracy}
\begin{figure}[!t]
 \centering
 \includegraphics[width=0.36\textwidth]{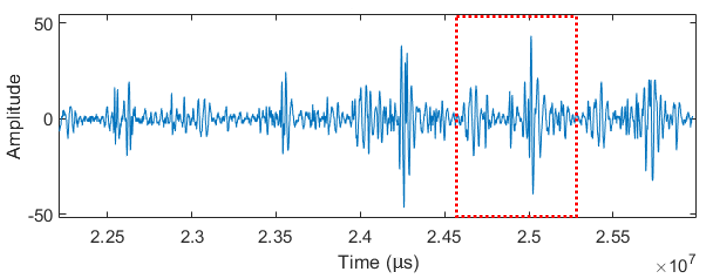}
 \caption{\textbf{Isolation of an elephant's seismic event} }
 \label{fig:seimicevent}
\end{figure}

To validate the accuracy of event detection, an aggregated dataset was created using isolated seismic events of elephant footfalls from Experiment C, and human footfalls and motorcycle data from Experiment A. When creating the dataset, a reference window of 772 ms (see Fig. \ref{fig:seimicevent}) was used to extract the individual footfall events from the seismic signals of Experiment C. The footfalls were correlated with the video footage (the ground truth) for their accuracy. A similar process was used to extract the human footfall events and seismic signals originating from a motorcycle from Experiment A. 

\begin{figure}[!t]
 \centering
 \includegraphics[width=0.4\textwidth]{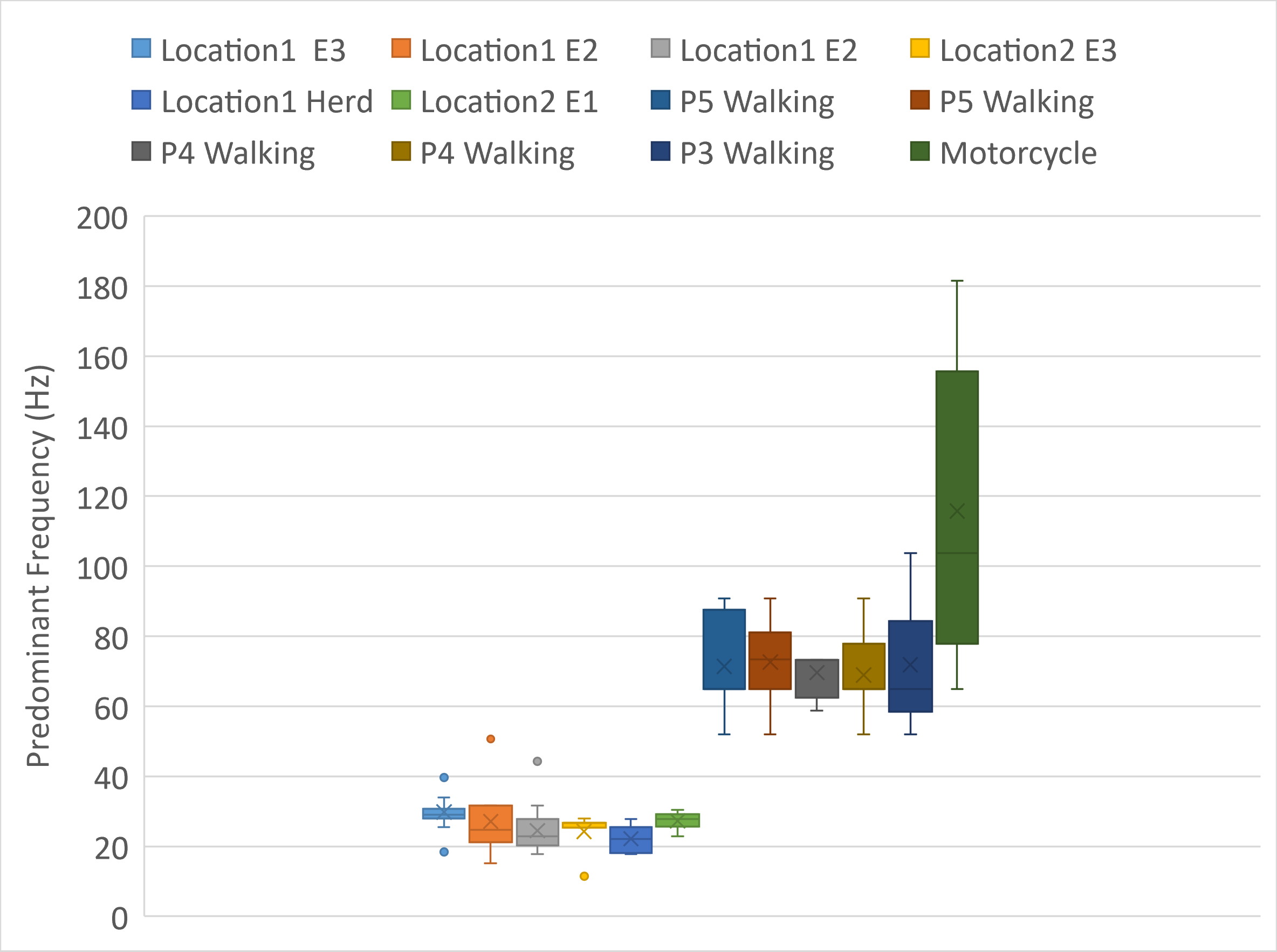}
 \caption{\textbf{Distribution of predominant frequency }}
 \label{fig:boxplot}
\end{figure}

Then predominant frequencies were used to identify the system's accuracy in elephant footfall identification. The obtained results are presented in Fig. \ref{fig:boxplot} in addition to the summary of predominant frequencies presented in Table \ref{tab:eventsummery}. In Fig. \ref{fig:boxplot}, `Location1 E3', `Location1 E2',` 'Location2 E3\footnote{scenario with the maximum detection range},' `Location1 Herd,' and `Location2 E1' represent the footfalls of the elephants in Experiment C. Similarly, P3, P4, and P5 denote the footstep signals of human subjects in Experiment A, and `Motorcycle' denotes the signal observed for the motorcycle. 

According to the figure, the predominant frequencies of elephant footfalls exhibits a visible clustering at 25 Hz. Additionally, a clear separation between the predominant frequencies of elephant and human footfalls can be observed when referring to Table \ref{tab:eventsummery}. Moreover, motorcycle events display a relatively \hl{wider spread of predominant frequencies} due to their noise-like patterns. A close analysis of the average predominant frequency reveals a slight deviation from 20 Hz, as mentioned in \cite{o2000seismic}, for the Asian elephant. The authors reckon that this is because the test subjects are Sri Lankan species (Elephas Maximus Maximus), which is a distinctive sub-species \hl{of} Asian elephants \hl{that} have distinctive physical characteristics. However, this assumption cannot be further discussed or proven due to the limited \hl{literature available}, and also due to the limited number of elephants available for the experiments.

A decision tree model with a 5-fold cross-validation was performed to classify the locomotion tasks by further annotating the dataset into two classes: 'Elephant' and 'Other' (encompassing human footfall and motorcycle events). The dataset contained 216 samples, including 67 Elephant footfalls and 149 Other activities. The model showed an average accuracy of 99.5\% in identifying elephant footfalls. The confusion matrix is provided in Table \mbox{\ref{tab:confusion_matrix}}.

\begin{table}[!t]
 \centering
 \caption{\textbf{Summery of predominant frequency distribution (event base)}}
 \begin{tabular}{|p{2cm}|p{2.5cm}|}
 \hline
 Scenario & Average Predominant Frequency\\
 \hline
 Elephant & 26.13 ($\pm$ 6.43) \\
 Human & 70.90 ($\pm$ 12.52) \\
 Motorcycle & 115.76 ($\pm$ 41.53) \\
 \hline
 \end{tabular}
 \label{tab:eventsummery}
\end{table}

\begin{table}[!t]
  \centering
  \caption{\textbf{Confusion matrix}}
\noindent
\renewcommand\arraystretch{1.5}
\setlength\tabcolsep{2.5pt}
\begin{tabular}{c >{\bfseries}r @{\hspace{0.2em}}c @{\hspace{0.0em}}c l}
 \multirow{10}{*}{\rotatebox{90}{\parbox{2.1cm}{\bfseries\centering True Class}}} & 
  & \multicolumn{2}{c}{\bfseries Predicted Class} & \\
 & & \bfseries Elephant & \bfseries Other \\
 & Elephant & \MyBox{65 }{ } & \MyBox{1}{} \\[2.4em]
 \cline{3-4}
 & Other & \MyBox{0}{} & \MyBox{150}{}\\
\end{tabular}
  \label{tab:confusion_matrix}
\end{table}

\section{Conclusion}
This study introduced a geophone-sensor interface to connect a geophone with an embedded system, with the motivation of non-invasive long-range elephant detection using seismic waves as a potential solution for human-elephant conflict. The geophone-sensor interface was tested under laboratory conditions and also in real-world scenarios for tamed, partly tamed, and untamed elephants. The system successfully captured seismic signals generated by elephant locomotion at a distance of 155.6 with an accuracy of 99.5\%.\hl{ The system was tested under different gain settings to study the seismic signals of elephant footfalls, ensuring the dynamic detection range supported by the geophone-sensor interface.} The system's stability was assessed by testing it over a frequency range from 1 Hz to 1 kHz and a temperature range from 10°C to 40°C. The results indicated only minor variations in magnitude and phase, even in challenging terrains with different temperatures, humidity levels, and soil conditions. Despite the system being tested only for Sri Lankan Elephants, due to the unavailability of African elephants, this will be a valuable tool for researchers interested in the seismic signal analysis of elephants in both Asian and African regions.

The future work of this study has two parts: the first one is to improve the ML model for real-time detection and classification of untamed elephants in their natural habitats \hl{ and the second will be to improve the system to automatically adjust the gain settings based on the saturation of the signal patterns.}

\bibliographystyle{unsrt}
\begin{CJK*}{UTF8}{gbsn}
\bibliography{bibliography}
\end{CJK*}
\EOD
\end{document}